\documentclass
[12pt]
{article}
\usepackage{latexsym,amsfonts}
\makeatletter
\@addtoreset{equation}{section}
\makeatother

\def\be{\begin{equation}}
\def\ee{\end{equation}}
\def\bea{\begin{eqnarray}}
\def\eea{\end{eqnarray}}
\def\bed{\begin{displaymath}}
\def\eed{\end{displaymath}}
\def\nn{\nonumber}

\def\a{\alpha}
\def\b{\beta}

\def\d{\partial}
\def\de{\delta}
\def\e{\epsilon}           % Also, \varepsilon

\def\f{\phi}               %      \varphi
\def\g{\gamma}
\def\h{\eta}

\def\j{\psi}

\def\l{\lambda}

\def\o{\omega}
\def\r{\rho}                      %     \varrho
\def\s{\sigma}                    %     \varsigma
\def\t{\tau}
\def\u{\upsilon}

\def\F{\Phi}
\def\G{\Gamma}

\def\L{\Lambda}

\def\S{\Sigma}
\def\U{\Upsilon}

\def\calD{{{\cal L}_U}}
\def\ce{{\cal E}}

\def\half{\frac{1}{2}}
\def\del{\partial}

\def\widebar{\overline}
\def\ul{\underline}

\def\divrt{\frac{1}{\sqrt{2}}}

\def\Rb{\widebar{R}}

\def\sb{\bar{\sigma}}
\def\cb{\widebar{\chi}}
\def\fb{\widebar{\phi}}               %      \varphi
\def\jb{\widebar{\psi}}

\def\ub{\widebar{\upsilon}}

\def\Ub{\widebar{\Upsilon}}
\def\da{{\dot \alpha}}
\def\db{{\dot \beta}}
\def\dg{{\dot \gamma}}

\def\et{{\tilde \e}}

\catcode`\@=11

\def\ppnumber#1{\gdef\@ppnumber{#1}}
\gdef\@ppnumber{\today}

\def\@maketitle{%
  \newpage
  \null
  \begin{flushright}
    \@ppnumber
  \end{flushright}
  \vskip 2em%
  \begin{center}%
  \let \footnote \thanks
    {\LARGE \@title \par}%
    \vskip 1.5em%
    {\large
      \lineskip .5em%
      \begin{tabular}[t]{c}%
        \@author
      \end{tabular}\par}%
    \vskip 1em%
    {\large \@date}%
  \end{center}%
  \par
  \vskip 1.5em}
\catcode`\@=12

\begin{document}

\title{{\bf Conformal Field Theory in Conformal Space}}

\author{
C.R. Preitschopf\\
Institut f\"ur Physik, Humboldt-Universit\"at zu Berlin\\
Invalidenstra\ss e 110, D-10115 Berlin, Germany\\
\and
M.A. Vasiliev\\
I.E. Tamm Theoretical Department, Lebedev Physical Institute\\
Leninsky Prospekt 53, 117924 Moscow, Russia}

\date{December 12, 1998}

\ppnumber{hep-th/9812113 \\ HUB EP-98/72\\ FIAN TD/98-23}

\maketitle

\begin{abstract}
  We present a new framework for a Lagrangian description of conformal
  field theories in various dimensions based on a local version of
  $d+2$-dimensional conformal space. The results include a true gauge
  theory of conformal gravity in $d=(1,3)$ and any standard matter
  coupled to it. An important feature is the automatic derivation
  of the conformal gravity constraints, which are necessary for the
  analysis of the matter systems.
  \vskip 0.5cm
  

  \noindent
  Keyword(s): local conformal symmetry, conformal field theory, 
  conformal gravity
\end{abstract}
\section{Introduction}

The concept of conformal space \cite{klein,weyl,kastrup1} was used by Dirac
\cite{dirac} to write the field equations for spinor and
Maxwell-fields in $d=(1,3)$ dimensional spacetime in manifestly
$SO(2,4)$-invariant form. He embedded Minkowski space as the
hypersurface $y^2=0$ in ${\mathbb{R}}{\rm P}^5$ and extended the
fields by homogeneity requirements to the whole of ${\mathbb{R}}^6$,
the space of homogeneous coordinates. This approach to conformal
symmetry proved quite useful \cite{kastrup2,macksalam,adler,ansou}, and
was employed frequently in the pre-string heydays of conformal field
theory \cite{cftreviews}. Only much later was this approach taken up
by Marnelius and Nilsson in the study of conformally invariant
particle mechanics \cite{mn1,mn2}, and Siegel was able to show
\cite{siegel} with conformal space techniques that one may describe
all free conformal fields in all dimensions in conformal space by a
simple and elegant particle mechanics \cite{hppt}.

Conformal gravity and conformal supergravity was studied extensively
in the context of gauged spacetime algebras
\cite{ktn1,ktn2,ktn3}, where the conformal group acts on a
fibre over a $d=(1,3)$ base space. If one intends to obtain ordinary
conformal gravity in such a framework, one has to impose constraints
on certain curvatures. These constraints are physically well
motivated, but they are imposed by hand. Nothing in the formalism
requires them. On the contrary, they explicitly break the original
local conformal symmetry, but physically equivalent symmetry
transformations can be obtained with the help of compensating
reparametrizations \cite{fradtseyt,pvncgrev}. Einstein gravity and
supergravity was of course also formulated as a gauge theory
\cite{chamwest,macdman}, with similar properties. In those cases,
however, there are also actions available with all symmetries manifest
and linearly realized \cite{chamseddine,stellewest,preitvas}. They are
constructed with the help of compensator fields and describe the
theory completely, without specification of additional constraints
beyond those arising as nondynamical equations of motion. 

We will extend this compensator framework to conformal gravity and
supergravity formulated in conformal space, which will serve 
as base manifold and in some sense also as fibre. Local $SO(2,4)$-gauge 
symmetry and 6-dimensional reparametrization invariance will be manifest. 

In the recent past there has been a number of studies of theories in
$d+2$ dimensions \cite{nishino,kounnas,bars1} which leave the 
framework of the original treatment of conformal space
\cite{cftreviews,mn1,mn2,siegel}. The field theory examples presented
in \cite{nishino} show similarities to second quantized fields in
conformal space, but not both extra null directions are removed. In
fact, the authors consider that feature one of the main points of their
theory: it is not a conventional $d$-dimensional theory ``in
disguise''. The same is true for the theories with two times of
\cite{kounnas,bars1}. More recently, in \cite{bars2}, the properties
of conformal space were gradually emphasized, and various gauge
choices were studied.

In this paper we do not construct a theory with 2 times. One of
the timelike directions in conformal space is removed by appropriate
gauge symmetries, and physical spacetime has the standard Lorentzian
structure. Our main goal is the construction of classical actions for 
second-quantized field theories. 

Our approach is similar to the group-manifold approach to conformal
supergravity \cite{cfn} in that the physical base space is embedded as
a hypersurface in our base manifold, and in most cases the action is
affine, i.e. it is written in terms of differential forms without
using a metric on the base manifold. However, the dimension of our
base manifold is typically that of the vector representation and hence
much smaller than the dimension of the gauge symmetry group.
Furthermore, our action is manifestly invariant under the whole 
gauge symmetry, not just some subgroup. 

In section~\ref{sec:confspace} of this paper we will set up our basic
formalism and conventions, and we construct conformal gravity in this
framework in section~\ref{sec:confgrav}. By judicious gauge fixing one
may obtain the usual form of conformal gravity. This we describe in
detail. Scalar fields are added in section~\ref{sec:scalars}, which
allows us to describe also Poincar\'e gravity. Subsequently we discuss
fermions, vector fields and gravitinos. We give a detailed account
of the gauge fixing procedure necessary to obtain the standard actions. 
The appendix summarizes our notation and conventions.

\section{Conformal Space}\label{sec:confspace}

We define $SO(2,d)$ gauge theory in $D=d+2$-dimensional spacetime
with coordinates $y^{\ul M}$ as follows:
for a $SO(2,d)$ vector $\F^N$ the covariant derivative is given
by
\be
D_{\ul M} \F^N \ = \
\partial_{\ul M} \ \F^{N} \ + \ \o_{\ul M}{}^N{}_K\F^{K} \ , 
\ee
and the curvature tensor $R_{\ul M \ul N}{}^{KL}$ by 
\be
\left[ D_{\ul M}, D_{\ul N} \right] \ \F^{K}  \ 
= \ R_{\ul M \ul N}{}^K{}_L \F^{L} \ .
\ee
Here ${\ul M,\ul N}\in \{0,1,\cdots,d-1,d+1,d+2\}$ are base space indices and 
$K,L,M,N \in \{0,1,\cdots,d-1,d+1,d+2\}$ denote vector indices of $SO(2,d)$.
The $SO(2,d)$-covariant Lie-derivative with respect to a vector
field $W^{\ul M}$ reads
\be
{\cal L}_{W} \ \F_{\ul N} \ \equiv \ W^{\ul M} \partial_{\ul M} \ \F_{\ul N}
                                \ + \  \partial_{\ul N} W^{\ul M} \F_{\ul M}
\ee
when acting on a base-space one-form, $SO(2,d)$ scalar $ \F_{\ul N}$,
\be
{\cal L}_{W} \ \F^{\ul N} \ \equiv \ W^{\ul M} \partial_{\ul M} \ \F^{\ul N}
                                \ - \  \partial_{\ul M} W^{\ul N} \F^{\ul M}
\ee
when $\F^{\ul N}$ is a base-space vector, $SO(2,d)$ scalar and
\be
{\cal L}_{W} \ \F_{N} \ \equiv \ W^{\ul M} \partial_{\ul M} \ \F_{N}
                                \ + \  W^{\ul M} \o_{\ul M}{}_N{}^K\F_{K}
\ee
when we differentiate a base-space scalar and a $SO(2,d)$
vector $\F_{N}$.

\subsection{Local Conformal Space}\label{sec:locs}

The equation $y^{\ul M} y_{\ul M}=0$ is incompatible with D-dimensional
reparametrization invariance, so we define a field $U^N(y)$ and
demand that effectively  
\be
U_M U^M \approx 0 
\ee
be satisfied. We will discuss below the precise implementation of this
constraint. 
The frame field $E_{\ul M}{}^N$ is given by
\be
E_{\ul M}{}^N \ = \ D_{\ul M}\  U^N
\ee
and we will assume that it is invertible, with inverse $E_N{}^{\ul M}$.
This formula is analogous to the definition of the frame field in the
context of AdS (super)gravity given in \cite{chamseddine,stellewest}.
We now contruct a $D$-dimensional metric
\be
\label{sixmetr}
G_{{\ul M} {\ul N}}\ =\ E_{{\ul M}}{}^K E_{{\ul N} K}\ ,\qquad
G^{{\ul M} {\ul N}}\ =\ E_K{}^{{\ul M}} E^{K {\ul N}}\ ,
\ee
with signature $(2,d)$. 

We use the base vector field $U^{\ul M} \equiv U^N E_N{}^{\ul M}{}$ to
ensure a projectivity condition that is a local version of the 
scaling condition $y^{\ul M} \partial_{\ul M} = h$ in global conformal
space \'a la Dirac. The theory should be independent of
the direction $U^{\ul M}$, and we realize that by demanding that our
fields be homogeneous in $U^{\ul M}$.  In the following, we will say
that a field $\F$ has scaling dimension $h$ if
\be
\calD \F \ = \ h \F .
\ee
Then $\partial_{\ul M} \F$ also has dimension $h$, since the Lie derivative
commutes with exterior differentiation,
%: $\calD d \F \ = \ d \calD \F $
and therefore
the covariant derivative $D_{\ul M}$ must carry dimension $h=0$.
Naively this would mean $h=0$ for any gauge connection $A_{\ul M}$,
but we cannot simply set $\calD A_{\ul M} =0$, since this breaks gauge
invariance. The correct condition arises from demanding that the
gauge convariant differential commute with the gauge covariant Lie-derivative:
\be
\label{LD}
\calD D \F \ - \ D \calD \F \ = \ (i_U D + D i_U) D \F - D i_U D \F
\ = \ i_U DD \F = \half \ i_U F \F ,
\ee
where we use the identity $\calD = ( i_U  D + D  i_U )$
with  $i_U H = p \ dy^{\ul N_1} \cdots dy^{\ul N_{p-1}}$ $U^{\ul M}
H_{\ul M \ul N_1 \cdots \ul N_{p-1}}$ for a $p$-form $H$.
This means that any curvature $F_{\ul M \ul N}$ will be required to 
satisfy the transversality condition
\be
\label{transf}
U^{\ul M} F_{\ul M \ul N} \ = \ 0 \ . 
\ee
In particular, this implies in the gravitational sector:
\be
U^{\ul M} R_{\ul M \ul N}{}^{KL} = 0 \ .
\label{transcurv}
\ee
Due to transversality and the Bianchi identities, 
the curvatures $F_{\ul M \ul N}$ have scaling dimension $0$. 
We emphasize that the gauge parameters do not obey any scaling condition.
This will allow us to choose the gauge $U^{\ul M}A_{\ul M}=0$, which
then leads to $\calD A_{\ul M} =0$ as well as $h=0$ for the residual gauge
parameters.
 
For the vector field $U^N$ we obtain $h=1$, i.e.
\be
\label{uweight}
\calD U^N \ = \ U^N \ ,
\ee
since
\be
\calD U^N \ = \ U^{\ul M} D_{\ul M} U^N \ = \  U^{\ul M} E_{\ul M}{}^N \
= \ U^N \ .
\ee
The frame field $D_{\ul M} U^M$ should therefore carry scale weight
$h=1$ as a consequence of (\ref{LD}), (\ref{transcurv}) and
(\ref{uweight}), and this is easy to verify explicitly:
\bea
\calD E_{\ul K}{}^N & = &   ( \partial_{\ul K}  U^{\ul M} )  E_{\ul M}{}^N
                         + U^{\ul M} D_{\ul M} E_{\ul K}{}^N \nn\\
& = &  D_{\ul K} U^N -  U^{\ul M} D_{\ul K} E_{\ul M}{}^N
                      + U^{\ul M} D_{\ul M} E_{\ul K}{}^N \nn\\
& = &  E_{\ul K}{}^N  +  U^{\ul M} R_{\ul M \ul N}{}^{N}{}_L U^L \nn\\
& = &  E_{\ul K}{}^N \ .
\eea
The covariant derivatives $D_N = E_N{}^{\ul M} D_{\ul M}$
satisfy the commutation relations:
\be
[ D_M , D_N ]\ \F^K \ = \ R_{MN}{}^{KL} \ \F_L - T_{MN}{}^L D_L \ \F^K
\ee
and we identify the torsion tensor as a particular set of components
of the gravitational curvature tensor: 
\be
T_{MN}{}^K \ = \  E_M{}^{\ul M}  E_N{}^{\ul N} R_{\ul M \ul N}{}^{K}{}_L 
U^L \ .
\ee
We have now ensured that the dependence of all our fields 
on the coordinate along the integral curves of the conformal Killing vector
field $U^{\ul M}$ is determined up to gauge transformations. 
The fields are specified by their values on a
hypersurface of codimension 1 in $D=d+2$-dimensional base space which
intersects the integral curves precisely once. Both this circumstance as
well as the constraint $U^M U_M\approx 0$ need to be 
incorporated properly into an action. 

\subsection{Global Conformal Symmetry}\label{sec:globconf}

Conformally flat spacetimes are characterized by 
\be
\label{zerocurv}
R^{MN} \ = \ 0 \quad ,
\ee
i.e. by a connection $\o^{MN}$ that is pure gauge. Of course,
it may not always be desirable to gauge away $\o^{MN}$, and therefore
we seek a general description of the symmetries of such spacetimes.
Gauge transformations that leave $\o^{MN}$ invariant obey
\be
\label{globaleps}
D \epsilon^{MN} \ = \ 0 \quad,
\ee
and this equation is integrable by~(\ref{zerocurv}) and admits
$\half (d+2)(d+1)$ independent solutions (one can fix arbitrary values
of $\epsilon^{MN}(y_0 )$ at some point $y_0^{\ul M}$ as integration
constants). The $\half (d+2)(d+1)$ conformal Killing vector fields
$\xi^{\ul N}$ are then defined by
\be
\label{killingu}
\xi^{\ul N} \ D_{\ul N} \ U^M  \ = \ \xi^{M} 
\ = \ - \ \epsilon^{MN} U_N \quad .
\ee 
They leave the metric $G_{\ul M \ul N}$ invariant:
\bea
\label{killinge}
{\cal L}_{\xi} \ E_{\ul M}{}^{M} & = & 
\xi^{\ul K} \ D_{\ul K} \ E_{\ul M}{}^M  
\ + \  \left( D_{\ul M} \xi^{\ul K} \right) \ E_{\ul K}{}^M 
\nn\\ 
& = & \xi^{\ul K} \ R_{\ul K \ul M}{}^{MN} U_N \ + \ D_{\ul M}\ \xi^{M}
\nn\\ 
& = & -\ \epsilon^{M}{}_{N} \  E_{\ul M}{}^{N} \quad,
\eea
which implies 
\be
\label{killingg}
{\cal L}_{\xi} G_{\ul M \ul N} = 0 \quad.
\ee
We interprete~(\ref{killingu})
as an invariance of $U^M$
under a combined gauge transformation
with covariantly constant parameter $\epsilon^{MN}$ and a reparametrization
with parameter $\xi^{\ul M}$.
Along with (\ref{killinge}) and~(\ref{killingg})
this implies invariance of the geometry under these
combined transformations which therefore are identified with
global conformal transformations. 
In order to show the equivalence of this presentation of global
conformal symmetry to a more standard description, we
pick the gauge $\o^{MN}=0$ and the parametrization $U^M(y) = y^M$,
so that we are in global conformal space. Now it is obvious that we
obtain just the ordinary $SO(2,d)$ - rotations of the coordinates
from the reparametrizations and the appropriate rotations of
$SO(2,d)$ - indices from the gauge transformations.

The vacuum solution (\ref{zerocurv})
of any theory invariant under diffeomorphisms and conformal
gauge transformations has global conformal symmetry
$SO(2,d)$ provided that the vacuum expectation values of all other
dynamical variables are invariant (vanish, for example). If the latter
property is not satisfied conformal symmetry is spontaneously broken 
and the fields carrying non-invariant vacuum expectation values 
are often called compensators. Note that from this
perspective the field $U^N$ is not a compensator. It plays a very
special role linking diffeomorphisms in the base manifold with 
gauge transformations in the fibre by (\ref{killingu}).

\subsection{Conformal Actions}\label{sec:confact}

The general form of the action principle we will use in this paper reads
\be
\label{genact1}
S \ = \   \int_{M_D} L \  \delta(U^2)
 \ \delta(\Psi)\ U^{\ul M} \partial_{\ul M} \Psi  \ .
\ee
$L$ is a gauge invariant $D$-form of conformal dimension 2, which means
\be
\label{lscaling}
\calD L\ = \ 2 L \ 
\ee
and ensures
\be
\partial_{\ul M} \left( \ U^{\ul M}\ L \ \delta(U^2) \ \right) = 0 \ .
\label{scaling}
\ee
This implies that the Lagrangian is independent of the coordinate
corresponding to the integral curves of the conformal Killing vector
field $U^{\ul M}$. The term $\delta(\Psi)$ eliminates a possible
volume divergence that may arise from integrating over this
coordinate.  In other words, $\Psi$ fixes a slice of $D$-dimensional
spacetime that should intersect each of the integral curves once. We
will call it a slice fixing condition. The term $U^{\ul M} \partial_{\ul M}
\Psi$ may be recognized as the Faddeev-Popov determinant for
reparametrizations of the orbit of the abelian group generated by the
vector field $U^{\ul M}$. We include it because 
it guarantees that the action is independent of any particular
slice choice determined by $\Psi$: 
under a local variation $\Psi \rightarrow \Psi + \Delta \Psi$ we get 
\bea
\delta S & = & \int_{M_D} L \  \delta(U^2)
 \ U^{\ul M} \partial_{\ul M} \left[\ \delta(\Psi)\ \Delta  \Psi \ \right]
\nn\\
 & = & \int_{M_D} \ \partial_{\ul M} \left[\ U^{\ul M} L \ \delta(U^2)
\ \delta(\Psi)\ \Delta  \Psi \ \right]
\nn\\
 & = & 0 \ .
\eea
The term $\delta(U^2)$ is a local version of the condition
$y^2=0$ on the projective coordinates of global conformal space.
As we shall see, it eliminates one more coordinate and we will
be left with an integral over a $d$-dimensional hypersurface 
embedded in $d+2$-dimensional conformal space. We note that we may 
rewrite~(\ref{genact1}) in the following form:
\be
\label{genact2}
S \ = \ \int_{M_D}\ L\ \delta(U^2)\ \delta(\Psi) 
\ i_U d \Psi  \ = \ (-)^{D+1}\int_{M_D}\ i_U L\ \delta(U^2)\ \delta(\Psi) 
 \wedge d \Psi  \ .
\ee
The equivalence is due to the fact that
$L \wedge d\Psi \equiv 0$ is a $D+1$ form in $D$ dimensions and therefore
$i_U (L \wedge d\Psi)$ also vanishes.

The above action is invariant under $D$-dimensional diffeomorphisms, 
$SO(2,d)$ gauge transformations and arbitrary local variations of $\Psi$.

\subsection{Generalizations}\label{sec:topact}

An action of the type
\be
\label{genact3}
S \ = \ \int_{M_D} L^{(D-1)} \ \delta(U^2) \ \delta(\Psi) \wedge  d \Psi  \
\ee
with a Lagrangian $(D-1)$ - form
$ L^{(D-1)}$ is $\Psi$-independent iff
\be
\label{ts}
d \Big( L^{(D-1)} \ \delta(U^2) \Big)=0\,,
\ee
which is equivalent, for nonvanishing $U^{\ul M}$, to
\be
\label{gencond3}
i_U \ \Big[ \ d \wedge \Big( L^{(D-1)} \ \delta(U^2) \Big) \ \Big] 
\ = \ \calD  \Big( L^{(D-1)} \ \delta(U^2) \Big) 
\ - \ d \wedge \Big( i_U L^{(D-1)} \ \delta(U^2) \Big)
\ = \ 0 \ .
\ee
The equivalence is due to the fact that in D dimensions
\be
\label{id}
V^{\ul M} A_{[\ul M \ul M_2 \ldots \ul M_D]} = 0 \quad
\Longrightarrow \quad A_{[\ul M_1 \ul M_2 \ldots \ul M_D]} = 0 \
\label{formscale}
\ee
if the vector field $V^{\ul M}$ is nonvanishing.

So, in general our Lagrangian $L^{(D-1)}$ has to obey a scaling
condition up to a total derivative term. In this paper
we demand most of the time strict scaling, i.e. 
\be
\calD  \Big( L^{(D-1)} \ \delta(U^2) \Big)=0 \ .
\ee
Then we are left with the condition 
$d \wedge \Big( i_U L^{(D-1)} \ \delta(U^2) \Big)=0$.
In the action~(\ref{genact1}) we solve it as follows:
the Lagrangian $L$ is related to $L^{(D-1)}$ by
\be
(-)^{D-1} \ L^{(D-1)} \ = \ i_U L \ ,
\ee
which immediately implies $i_U L^{(D-1)} = 0$. The reverse procedure,
constructing $L$ from a given $L^{(D-1)}$ with $i_U L^{(D-1)}=0$,
requires solving a nontrivial cohomology problem: we can always
add a term $L^\prime$ to $L^{(D-1)}$ such that $i_U L^\prime$ 
is proportional to $U^2$. 

It is worth mentioning that since the $\Psi$ - independence condition
requires (\ref{ts}), the Lagrangian form $L^{(D-1)}\delta (U^2)$ is
closed. In other words our action is in a certain sense topological.
This is of course expected because the dynamics is located on the
boundary singled out by the slice fixing condition and is required to
be independent of a particular slice choice.  We thus arrive at a
standard cohomology problem: exact forms
$L^{(D-1)}\delta(U^2)=dl^{(D-2)}$ with local functionals $l^{(D-2)}$
give rise to trivial equations of motion. A typical total derivative 
in physical spacetime is written in conformal space as
\be
\label{totd}
d \ H\ \delta(U^2) \wedge dU^2\ \delta(\Psi) \wedge d \Psi \ ,
\ee
with some $d-1$-form $H$.

An important example is provided by topological or Chern-Simons
actions, which have the form
\be
\label{topact} 
S_{top} \ = \ \int_{M_D} L_{top} 
\ \delta(U^2) \wedge  dU^2\ \delta(\Psi) \wedge d \Psi \ .
\ee
$L_{top}$ is a $(D-2)$ - form that satisfies
\be
\label{condtop}
i_U \left(\ d \ L_{top} \ \right) \ = \ 0 \ .
\ee 
Note that in this case we will generally not have strict scaling.
If $L_{top}$ is constructed as a wedge product of curvatures,
we do: $i_U L_{top}=0$ by~(\ref{transf}),
$L^{(D-1)} = L_{top}\wedge dU^2$ and therefore
$i_U L^{(D-1)} = 2 \ U^2 L_{top}$ by~(\ref{uweight}).
Specific examples of such actions will be given below.

\section{Conformal Gravity}\label{sec:confgrav}

Conformal gravity in $d=(1,3)$ is described by the action
\be
\label{sixact}
S \ = \ - \frac{1}{8}\  \int_{M_6}
\ \epsilon_{N_1 \ldots N_6} \ E^{[N_1} \wedge E^{N_2} \wedge R^{N_3 N_4}
\wedge R^{N_5 N_6]} \ \delta(U^2)
\ \delta(\Psi) \ U^{\ul M} \partial_{\ul M} \Psi \ .
\ee
The Lagrangian obviously has the requisite scaling property~(\ref{lscaling}). 
Apart from the fact that it is an invariant of global symmetries,  
the field $U^M$ plays a role reminiscent of the 
compensator in ordinary gravity \cite{chamseddine,stellewest,preitvas}.
This is not surprising, since the action~(\ref{sixact}) has more local
symmetries than we would expect of conformal gravity. In fact we will
show that we can choose a gauge for $U^M$ such that we are
left with the usual $R^2$-action of conformal gravity, along with the
standard curvature constraints \cite{ktn1}. For conformally flat 
gravitational fields, it may be more useful to gauge away the connections
instead, and the geometry is then encoded entirely in $U^M$.

Let us now explain a particular way of partial gauge fixing that
achieves the reduction to $d=4$. By a six-dimensional 
reparametrization we can set generically
\be
\label{y6}
U^{\ul N} \ = \ \delta^{\ul N}_{\ul \oplus} \ .
\ee
This leaves us with $y^{\ul \oplus}$~-independent
diffeomorphisms, generated by six-dimensional 
vector fields $\Xi^{\ul N}$ which satisfy:
\be
\partial_{\ul \oplus}\ \Xi^{\ul N} \ = \ 0 \ .
\ee
We partially fix $SO(2,4)$-gauge invariance by requiring
\be
\label{o+}
\omega_{\ul \oplus}{}^{MN} \ = \ 0 .
\ee
Further gauge transformations must then be generated by
$y^{\ul \oplus}$~-independent parameters:
\be
\partial_{\ul \oplus}\ \Lambda^{M N} \ = \ 0 \ .
\ee
As a consequence the $h=1$ frame field obeys
\be
E_{\ul \oplus}{}^N \ = \ \partial_{\ul \oplus}\ U^N \ = \ U^N \ ,
\ee
which implies
\be
U^N \ = \ {\rm e}^{y^{\ul \oplus}} \ V^N
\ee
for some $y^{\ul \oplus}$~-independent vector $V^N$,
\be
\partial_{\ul \oplus}\ V^N \ = \ 0 \ ,
\ee
as well as
\bea
E_{\ul \oplus}{}^N & = &  {\rm e}^{y^{\ul \oplus}} \ V^N \nn \\
E_{\ul \mu}{}^N & = &  {\rm e}^{y^{\ul \oplus}} \ D_{\ul \mu} V^N  \ ,
\ \ul \mu \in \{\ul \ominus, 0,1,2,3 \} \ .
\eea
In this gauge, transversality of the curvature~(\ref{transcurv}) 
is equivalent to 
\be
\partial_{\ul \oplus}\ \omega_{\ul \mu}{}^{MN} = 0 \  .
\ee
We have now determined the $y^{\ul \oplus}$~-dependence of each field
that appears in the action~(\ref{sixact}), and by scaling~(\ref{scaling}) 
the Lagrangian does not depend at all on $y^{\ul \oplus}$.
We now choose the slice condition
\be
\label{slice}
\Psi \ = \ y^{\ul \oplus} \ ,
\ee
and the partially gauge fixed action simplifies to
\be
S \ = \ - \frac{1}{4} \ \int_{M_5}
\ \epsilon_{N_1 \ldots N_6} \
V^{[N_1} E^{N_2} \wedge R^{N_3 N_4} \wedge R^{N_5 N_6]}
\ \delta(V^2) \ .
\label{fiveaction}
\ee
It is still manifestly $SO(2,4)$-gauge invariant and affine and therefore
reparametrization invariant, but now only in a 5 dimensional base space.
We introduce the notation
\be
f(y^{\ul \ominus}, x^{\ul m}) \big| \equiv f(0, x^{\ul m}) \ ,
\ee
and assume that
\bea
V^{\oplus} \Big| & \not = & 0 \nn \\
D_{\ul \ominus} V^\ominus \Big| & \not = & 0 \ , 
\eea
which is generically true.  
Then we may choose the gauge
\bea
V^{\ominus} \Big| & = & 0 \nn \\
V^{m} \Big| & = & 0 \ 
\eea
by an appropriate choice of $\Xi^{\ul \ominus} \big|$ 
and $\L^{m \ominus} \big|$.
An immediate consequence is
\be
\de(V^2) \ = \ \frac{1}{2 V^{\oplus} \ 
D_{\ul \ominus} V^{\ominus} } \ \de( y^{\ul \ominus} ) \ .
\ee
As we anticipated,
this term eliminates the coordinate $y^{\ul \ominus}$ from the
action~(\ref{fiveaction}). 
We now use the parameters $\partial_{\ul \ominus} \Xi^{\ul m} \big|$ to 
set $E_{\ul \ominus}{}^m \big| = 0$. The nonzero components of the frame field
are:
\bea
\label{fourframe}
E_{\ul \oplus}{}^\oplus \Big| &  = & \r(x) \ = \ V^\oplus \Big| \nn \\
E_{\ul \ominus}{}^\ominus \Big| &  = & \s(x) \ = \ D_{\ul \ominus} 
V^\ominus \Big| \nn  \\
E_{\ul \ominus}{}^\oplus \Big| &  = & \t(x) \nn \\
E_{\ul m}{}^\oplus \Big| &  = & \partial_{\ul m} \r(x) +
           \r(x) \ \o_{\ul m}{}^{\oplus \ominus} \Big| \nn \\
E_{\ul m}{}^n \Big| &  = & \r(x) \ e_{\ul m}{}^{n} \Big| \ ,
\eea
where we define
\be
\label{fr}
e_{\ul m}{}^{n }=\o_{\ul m}{}^{n \ominus}\,.
\ee
The nonvanishing components of the inverse frame field are
then
\bea
E_{\oplus}{}^{\ul \oplus} \Big| &  = & \r^{-1}(x) \nn\\
E_{ \ominus}{}^{\ul \oplus} \Big| &  = & - (\r \s)^{-1} \t(x) \nn \\
E_{ \ominus}{}^{\ul \ominus} \Big| &  = & \s^{-1}(x) \nn \\
E_m{}^{\ul \oplus} \Big| &  = & -\r^{-2}(x) \
e_m{}^{\ul n}\ \left(\partial_{\ul n} \r(x) +
                \r(x) \ \o_{\ul n}{}^{\oplus \ominus} \right) \Big| \nn\\
E_{m}{}^{\ul n} \Big| &  = & \r^{-1}(x) \  e_{m}{}^{\ul n} \Big| \ .
\label{infr}
\eea

We insert~(\ref{fourframe}) into~(\ref{fiveaction}) and obtain
\bea
\label{fouraction}
S & = & \frac{1}{8} \ \int_{M_4}
\epsilon_{m_1 \ldots m_4}
\ \Big[ R(M)^{m_1 m_2}  \wedge R(M)^{m_3 m_4} \nn\\ 
&& \phantom{ - \frac{1}{4} }
+ 8 \frac{\r}{\s} R_{\ul \ominus}{}^{m_1 \ominus}  \wedge e^{m_2}  
\wedge R(M)^{m_3 m_4}
\ - \ 8 \frac{\r}{\s} 
R_{\ul \ominus}{}^{m_1 m_2} \wedge  e^{m_3}  \wedge 
R(P)^{m_4} \Big] \ ,
\eea
where we use the decomposition
\be
\label{confcurv}
\begin{array}[c]{ccccl}
\half R(P)^{m} & \equiv & \half R^{m \ominus} & = &
     de^m + \o^{m}{}_{k} e^k + b e^m 
     \\
\half R(M)^{mn} & \equiv & \half R^{mn} & = &
     d\o^{mn} + \o^{m}{}_{k}\o^{kn} -2 e^{[m} f^{n]}
     \\
\half R(D) & \equiv & \half R^{\ominus \oplus} & = & db - e^m f_m 
     \\
\half R(K)^{m} & \equiv & \half R^{m \oplus} & = &
     df^m + \o^{m}{}_{k} f^k - b f^m  \quad .
\end{array}
\ee
The curvatures $R_{\ul \ominus}{}^{m_1 \ominus} =
dx^{\ul m} \partial_{\ul \ominus} \o_{\ul m}{}^{m_1 \ominus} + \cdots $
and
$R_{\ul \ominus}{}^{m_1 m_2} =
dx^{\ul m} \partial_{\ul \ominus} \o_{\ul m}{}^{m_1 m_2} + \cdots $
are independent one-forms in four-dimensional spacetime and therefore
play the role of Lagrange multipliers enforcing the constraints
\bea
R(P)_{\ul m \ul n}{}^{k} & = & 0 \nn \\
e_k{}^{\ul n} R(M)_{\ul m \ul n}{}^{k l} & = & 0 \ ,
\label{allcon}
\eea
where $e_k{}^{\ul n}$ is inverse of
$e_{\ul m}{}^n $. These constraints are the sole remnant of
the extra dimensions we started with. In the usual treatment
of conformal gravity they need to be put in by hand, whereas here
they follow from the action~(\ref{sixact}). For later reference we give their
invariant form:
\bea
dU^2 \wedge U^{[P} R^{M]N} U_N & = & 0 \nn \\
dU^2 \wedge E^{[M} \wedge R^{NP} U^{Q]}   & = & 0 \ .
\label{allconinv}
\eea

We are now left with the usual description of conformal gravity
\be
\label{ca}
S \ = \ \frac{1}{8}\ \int_{M_4}
\epsilon_{m_1 \ldots m_4}
R(M)^{m_1 m_2} \wedge R(M)^{m_3 m_4}\,,
\ee
where the conformal boost gauge fields $\omega_{\ul m}{}^{k \oplus} $
and the Lorentz gauge fields $\omega_{\ul m}{}^{kl}$ are expressed in
terms of the vierbein $e_{\ul m}{}^n$ and 
$b_{\ul m}=\o_{\ul m}{}^{\ominus \oplus}$
by virtue of the constraints~(\ref{allcon}).
They are solved in $d$ dimensions by
\bea
\label{allsol}
\o_{kmn} \equiv e_k{}^{\ul m} \o_{\ul m m n} & = &
-  e_{[n}{}^{\ul l}  e_{k]}{}^{\ul k} \left( \partial_{\ul k} + 
b_{\ul k} \right) e_{\ul l m}
-  e_{[k}{}^{\ul l}  e_{m]}{}^{\ul k} \left( \partial_{\ul k} + 
b_{\ul k} \right) e_{\ul l n}
\nn\\ && 
+  e_{[m}{}^{\ul l}  e_{n]}{}^{\ul k} \left( \partial_{\ul k} + 
b_{\ul k} \right) e_{\ul l k}
\nn\\ 
\o^m{}_{mn} & = & \frac{1}{e} \partial_{\ul k} \left(e_{n}{}^{\ul k} e\right)
\ + \ (d-1) b_n 
\nn\\
R(\o)_{\ul m \ul n}{}^{m n} & = &
R(e)_{\ul m \ul n}{}^{m n} - 4  e_{[\ul m}{}^{[m} D^{\cal L}_{\ul n]} b^{n]}
+ 4 e_{[\ul m}{}^{[m} b_{\ul n]} b^{n]} 
- 2 e_{[\ul m}{}^{m} e_{\ul n]}{}^{n} b^k b_k    
\nn\\
R(\o)_{\ul m}{}^{n} & = &
R(e)_{\ul m}{}^{n} + (d-2)  D^{\cal L}_{\ul m} b^{n} +  
e_{\ul m}{}^{n} D^{\cal L}_{k} b^{k}
 - (d-2) [ b_{\ul m} b^{n} - e_{\ul m}{}^{n}  b^k b_k ]
\nn\\
\o_{\ul m}{}^{m \oplus} \ \equiv \ f_{\ul m}{}^{m} & = & 
- \frac{1}{d-2} R(\o)_{\ul m n}{}^{n m} \ + 
\ \frac{1}{2(d-1)(d-2)} R(\o)_{kn}{}^{nk} e_{\ul m}{}^{m}
\nn\\
\o_{n}{}^{n \oplus} \ \equiv \ f_{\ul m}{}^{m} \ e_m{}^{\ul m}
& = & -\frac{R(e)}{2(d-1)}\  
- D^{\cal L}_m b^m - \frac{d-2}{2} b^m b_m \ ,
\eea
where $e$ is the determinant of the d-bein, $
D^{\cal L}_{\ul m}$ is the standard torsionless Lorentz connection:
\be
D^{\cal L}_{\ul n} A_n =\partial_{\ul n} A_n +\o(e)_{\ul n}{}_n{}^m A_m \ ,
\ee
$R(e)_{\ul m \ul n}{}^{m n}$ is the corresponding $SO(1,3)$-curvature, 
$R(e)_{\ul m}{}^{m}= R(e)_{\ul m n}{}^{n m}$ the Ricci-tensor and $R(e) =
R(e)_{m}{}^{m}$ the Ricci-scalar.  At this stage all indices are
raised and lowered with the d-bein $e_{\ul m}{}^m$. If we
insert~(\ref{allsol}) into $R(M)^{m n}$, we obtain the totally traceless
part of $R(\o)_{\ul m \ul n}{}^{m n}$, i.e. the Weyl-tensor:
\bea
R(M)_{\ul m \ul n}{}^{m n} & = &
R(\o)_{\ul m \ul n}{}^{m n} \ - \ 4 \ e_{[\ul m}{}^{[m} f_{\ul n]}{}^{n]}
\\ & = & 
R(e)_{\ul m \ul n}{}^{m n} \ + \ \frac{4}{d-2} 
e_{[\ul m}{}^{[m} R(e)_{\ul n]}{}^{n]} \ - \ \frac{2}{(d-1)(d-2)}        
e_{[\ul m}{}^{[m} e_{\ul n]}{}^{n]} R(e)\ .\nn
\eea
In this expression $b_{\ul m}$ of course does not appear anymore.
We can trace the absence of 
the dilatation gauge field $\omega_{\ul m}{}^{\ominus \oplus }$ 
in the action~(\ref{ca}) to the residual local gauge symmetry
\be
\label{dilga}
\delta
\o_{\ul n}{}^{\ominus \oplus} = - e_{\ul n}{}^n \Lambda_{n}{}^{\oplus} \Big|\ ,
\ee
which is a shift that 
we may use to set $\omega_{\ul m}{}^{ \ominus \oplus} = b_{\ul m} $ to zero.

The normalization of~(\ref{ca}) is chosen such that at the linearized level,
i.e. when $g_{\ul m\ul n} = \eta_{\ul m\ul n} + h_{\ul m\ul n}$, it yields
the standard higher derivative action:
\be
\label{calin}
S \ = \ - \frac{1}{4}\ \int_{M_d}
h_{\ul m\ul n} \pi^{\ul m\ul n}{}_{\ul r\ul s}
\Box \Box
h^{\ul r\ul s} \ + \ O(h^3)
\ee
with  $\pi^{\ul r\ul s}{}_{\ul m\ul n} = \pi^{(\ul r}{}_{\ul m} 
\pi^{\ul s)}{}_{\ul n} - \frac{1}{d-1} \pi^{\ul r\ul s} \pi_{\ul m\ul n}$
and $\pi_{\ul m\ul n} = \eta_{\ul m\ul n} - 
\Box^{-1}\partial_{\ul m}\partial_{\ul n}$.

In the context of the present paper, the most important output of
conformal gravity is the automatic derivation of the
constraints~(\ref{allcon}) which are necessary for the analysis of various
matter systems. 

\section{Scalars}\label{sec:scalars}

The fundamental field representation of the conformal group is the
scalar field. It allows a free field description in any dimension, and
only the spinor field shares that property. In this section we will
present the coupling of scalars to the gauge fields of the conformal
group, and in section \ref{sec:fermion} we repeat the exercise for spinors.
Since we want to realize conformal symmetry linearly on all fields,
and since scalars do transform under conformal transformations 
except in $d=2$, we prefer not to assign
scalar fields the trivial representation of $SO(2,d)$ for $d\neq 2$.  

\subsection{$d \not= 2$}

We describe a conformal scalar matter field in $d$ - dimensional
spacetime with $d \not= 2$ by a vector $\Phi^M$ of the d - dimensional
conformal group $SO(2,d)$. $d+1$ components of $\Phi^M$ will be 
identified with the physical field
\be
\varphi = U^M \Phi_M\ 
\label{physcalar}
\ee
and its space-time derivatives. The remaining component is eliminated
by requiring the Lagrangian to be invariant under the gauge transformation
\be
\label{gscal}
\delta \Phi^M = U^M \eta (y) \ .
\ee
The field $\Phi^M $ is defined to have the scaling dimension $h=-\frac{d}{2}$.
As a result the physical field $\varphi$
has dimension $-\frac{d}{2}+1$ while
$\eta (y)$ is an arbitrary parameter of dimension $-\frac{d}{2}-1$:
\be
\label{s}
\calD \Phi^M =-\frac{d}{2}\Phi^M \,,\qquad \calD \varphi
=-(\frac{d}{2}-1)\varphi \,, \qquad\calD \eta =-(\frac{d}{2}+1)\eta\,.
\ee

In addition we require a symmetry
\be
\label{restrict}
\delta \Phi^M \ = \ U^2 \ \Xi^M
\ee
with $h_\Xi = h_\Phi - 2$ and otherwise arbitrary $\Xi^M(y)$. This would
imply that our fields
contribute to the action only through their restriction to the
hypersurface $U^2 =0$. In other words, this symmetry guarantees
that in the coordinate choice of
section \ref{sec:confgrav} with
\be
0 \ = \ 
U^2 \ = \ 2 \r(x) \s(x) y^{\ul \ominus} \ + \ O(  (y^{\ul \ominus})^2 )
\ee
the components $\partial_{\ul \ominus} \Phi^M \big|$
which are independent fields in the physical $d$-dimensional
spacetime do not contribute. Since $\partial_{\ul \ominus}\Phi^M \big|$
serve as Lagrange multipliers the symmetry
(\ref{restrict}) guarantees the absence of d-dimensional constraints
on the fields $\Phi^M$. In the case under
consideration this is necessary because
the ensueing constraint would be too strong: it
enforces $\varphi =0$.
%This property will also be respected by interactions.
%For the conformal gravity action a symmetry of the type~(\ref{restrict}) is
%not appropriate, since it would remove the Lagrange multipliers that enforce
%the vital constraints~(\ref{allcon}).

The only first order action compatible with the symmetries
(\ref{gscal}) and (\ref{restrict}) is
\be
\label{scalaffact}
S = a \ \int_{M_{d+2}} |E|
\Bigg[
\ \Phi^M U_M D_N \Phi^N
\ - \ \Phi^M U_N D_M \Phi^N
\ - \ \frac{d}{2}\ \Phi_M \Phi^M
\Bigg]
\ \delta(U^2)
\ \delta(\Psi) \ U^{\ul M} D_{\ul M} \Psi \ ,
\ee
where
\be
|E|=\frac{1}{(d+2)!}
\epsilon_{N_1 \ldots N_{(d+2)}} \ E^{N_1} \wedge\ldots\wedge E^{N_{(d+2)}}\,.
\ee
It can be rewritten as
affine action (without inverse frame fields) by
\be
|E| \ D_M  \ = \ \frac{1}{(d+1)!}
\ \epsilon_{N_1 \ldots N_{(d+1)} M} \ E^{N_1} \wedge\ldots\wedge E^{N_{(d+1)}}
\wedge D \  \ .
\ee
The symmetry~(\ref{restrict}) is explicit since the differential
operator $U_M D_N -  U_N D_M$ commutes with $U^2$. Up to a total derivative,
i.e. a term of the type~(\ref{totd}), the bilinear form in the scalar 
fields in the action~(\ref{scalaffact}) is symmetric. This property is 
not obvious and is true only for fields $\Phi^M$ with the correct 
scaling dimension $h[\Phi^M]=-d/2$. 

In the Appendix we show that there exists a
two parameter class of second-order actions symmetric under
(\ref{gscal}) and (\ref{restrict}) which however are all equivalent
to (\ref{scalaffact}) by field redefinitions or modulo total
derivatives. The simplest action of this class is
\bea
\label{g}
S & = & \int_{M_{d+2}} |E|
\left[ \ \varphi \ D_M D^M \varphi \ \right]
\ \delta(U^2)
\ \delta(\Psi) \  U^{\ul M} \partial_{\ul M} \Psi\ ,
\eea
which can again be shown to be
symmetric in the scalars up to a total derivative. 
It is remarkable that the same physical system can, in $d \not= 2$, be
described in terms of completely different representations
of the conformal group: for (\ref{scalaffact})
we use the vector representation
$\Phi^M$, while for (\ref{g}) the scalar representation $\varphi$ suffices.
Note that the naive action
\bea
\label{gr}
S & = & \int_{M_{d+2}} |E|
\left[\  D_M \varphi  \   D^M \varphi \ \right]
\ \delta(U^2)
\ \delta(\Psi) \  U^{\ul M} \partial_{\ul M} \Psi
\eea
is nondynamical, and in fact completely trivial,
since it is not invariant under (\ref{restrict}).

We therefore will now show
that~(\ref{scalaffact}) describes a conformally coupled
scalar field in $d$ dimensions postponing a detailed account of the second
order actions
to the appendix.
To this end we use~(\ref{infr}) and
fix the gauge invariance~(\ref{gscal}) by setting
\be
\label{p+}
\Phi^\oplus =0\ .
\ee
When we take into account the gauge condition $\o_{{\ul \oplus}}{}^{MN}=0$,
this allows us to reduce~(\ref{scalaffact}) to the form:
\bea
 S  =  \frac{a}{2} \int_{M_d} \r^d |e|
\ \Bigg[
\ \r \Phi_\oplus  D_n \Phi^n
\ - \ \r \ \Phi^n   D_n \Phi_\oplus
\ - \  \frac{d}{2}\ \Phi_n \Phi^n \ \Bigg]\ ,
\eea
where 
\be
|e|=\frac{1}{d!}
\epsilon_{n_1 \ldots n_d} \ e^{n_1} \wedge\ldots\wedge e^{n_d}
\ee
and
\be
D_n \ = \ E_n{}^{\ul M} D_{\ul M}
\ = \ \frac{1}{\r} e_n{}^{\ul m} D_{\ul m} 
\ + \ E_n{}^{\ul \oplus} D_{\ul \oplus} \ .
\ee
The first term is the $SO(2,4)$-covariant derivative in $d=4$ dimensional
spacetime, and the second is an additional term reflecting the scaling
properties of our fields. 
Upon redefining
\be
\label{resc}
\Phi^n =\r^{-\frac{d}{2}}\phi^n \qquad,\qquad
\Phi_{\oplus} =\r^{-\frac{d}{2}}\phi \qquad,\qquad 
\varphi =\r^{-\frac{d}{2}+1}\phi
\ee
one finds
\bea
\label{re}
\r D_n \Phi_{\oplus} & = & \r^{-d/2} e_n{}^{\ul n}
\left( \partial_{\ul n} \phi - \frac{1}{2}(d-2) b_{\ul n} \phi \right)
 -  \r^{-d/2} \phi_n  
\nn\\
\r D_n \Phi^n & = &  \r^{-d/2} \left(
\frac{1}{e} \partial_{\ul n} \left( e e_n{}^{\ul n} \phi^n \right)
+ \frac{1}{2}(d-2) b_n \phi^n + \o_n{}^{n \oplus} \phi \right)
\ .
\eea
The action now reads
\bea
\label{a1}
S  =   \frac{1}{2} \int_{M_d} |e|
\ \Bigg\{  \o_n{}^{n \oplus} \ \phi^2
\ - \ 2 \ \phi^n
\ \left[
\partial_n \phi - \frac{1}{2}(d-2) b_n \phi + \frac{1}{4} (d-2) \phi_n
\right] \Bigg\} \ .
\eea
The fields $\phi^n$ are auxiliary and are expressed in terms of
derivatives of $\phi$ by means of their equations of motion
\be
\label{auxe}
\phi_{n} = - \frac{2}{d-2}\ \partial_{n} \phi \ + \ b_n \phi \, ,
\ee
which leads to the equivalent action
\bea
\label{a2}
S & = &  \frac{a}{d-2} \int_{M_d} |e|
\left[\ \partial_{n} \phi \ \partial^{n} \phi
\ - \ \frac{(d-2)}{4(d-1)}\, R(e)\, \phi^2
\ \right] \ .
\eea

Similar to the case of pure conformal gravity the dilatation gauge
field $\o_{\ul n}{}^{\ominus\oplus}$ does not appear in the final action.
The Ricci scalar arises due to~(\ref{allsol}). If we choose
$a= 1-\frac{d}{2}$,
we obtain the standard action for a conformally coupled massless scalar in an
external gravitational field. It possesses the local scale invariance \be
\label{locdil} \delta e_{\ul n}{}^n \ = \ \e(x) \, e_{\ul n}{}^n\qquad,\qquad
\delta \phi \ =\ -\frac{1}{2}(d-2)\ \e(x) \, \phi \ .  \ee
When one looks for the origin of this symmetry, one has to take into account the
definitions of the frame field~(\ref{fr}), of the scaling factor
$\r$~(\ref{fourframe}) and of the physical scalar $\phi$~(\ref{resc}).  Then one
may trace it, for the fixed $y^{\ul \oplus}$~-diffeomorphism gauge that we
described, to the local dilatation symmetries with parameter
$\Lambda^{\oplus\ominus}$.
Alternatively, one may fix dilatation symmetries and perform 
reparametrizations with $\Xi^{\ul \oplus} = y^{\ul \oplus} \e(x)$. 
Now we would claim that dilatations are a remnant of the extra dimensions
we introduced. Yet another way to interprete these dilatations 
is to fix reparametrizations and gauge transformations, and 
change the slice fixing function $\Psi$ appropriately. 

One can easily introduce conformal selfinteractions for scalars as
\be
\label{scalint}
S^{int}  =  2 \lambda \int_{M_{d+2}} |E|
\ \varphi^{\frac{2d}{d-2}}
\ \delta(U^2)
\ \delta(\Psi)\  U^{\ul M} \partial_{\ul M} \Psi \ ,
\ee
where $\lambda$ is an arbitrary real dimensionless coupling constant.
This action is invariant under the transformations~(\ref{gscal})
and (\ref{restrict})
because they imply $\delta \varphi = U^2 \eta$, which
yields zero when integrated with the above measure.
The power of the selfinteraction gives us precisely
the right scaling properties. We may reduce~(\ref{scalint}) to the
ordinary $d$ - dimensional action
\be
S^{int}  =  \lambda \int_{M_d} |e| \ \phi^{\frac{2d}{d-2}} \ .
\ee
For $d=4$ one arrives as expected at the standard $\phi^4$ interaction.

\subsection{$d=2$}

The above consideration is not immediately applicable to the
particular case of $d=2$ since some of the coefficients acquire
singularities at $d=2$. This is
because a 2-d massless scalar field is conformally invariant (cf.
e.g.~(\ref{s})) and therefore should be described by a singlet of the
conformal group $O(2,d)$ rather than by a vector as for $d \neq 2$.
Consequently the simplest action for a scalar field $\varphi$ is given
by
\be
\label{2scal}
 S \ = \ -\int_{M_{4}} |E|
\ G^{{\ul N} {\ul M}}
\ \d_{{\ul N} }\varphi
\ \d_{{\ul M} }\varphi
\ \delta(U^2) \ \delta(\Psi) \ U^{{\ul M}} \partial_{{\ul M}} \Psi\, ,
\ee
where $G^{{\ul N} {\ul M}}$ is the 
four-dimensional metric tensor~(\ref{sixmetr}).
Recall that for $d \not= 2$ this action is trivial: all fields are set
to zero by constraints. Now $\varphi$ has scale dimension $h=0$
\be
\calD \varphi = 0\ , 
\ee
and if we take into account~(\ref{infr}) it 
follows that~(\ref{2scal}) reduces to the
standard 2-d scalar action
\be
 S \ = \ -\frac{1}{2}\int_{M_{2}} |e| \ g^{{\ul n} {\ul m}}
\ \d_{\ul n} \varphi \ \d_{\ul m} \varphi\ .
\ee
Let us note that this action is a particular case of the action 
for $p$ - form fields considered in section \ref{sec:vector} below. 

\section{Compensators and Poincar\'e Gravity}\label{sec:poincare}

Compensators are fields that carry only pure gauge degrees of freedom.
They are used to describe physical systems in terms of variables
which increase manifest symmetries. The prime example is the formulation
of a massive vector boson in terms of a Higgs field and a $U(1)$-gauge field.
Compensators have been used extensively in the context of supergravity
\cite{kaktow,daskaktow,ferpvn,fergri,siegat,witholpro}
because they provide an organizing principle for the various auxiliary
fields that appear in off-shell supersymmetric actions
\cite{comprev}. The simplest 
compensator is a scalar field, and it may be used to describe Poincar\'e
gravity in conformally symmetric terms.

%\subsection{$d\neq 2$}

The action~(\ref{a2}) can be used in the compensator framework
provided that the field $\phi$ gets a non-vanishing expectation value.
Then one can use the local dilatation symmetry~(\ref{locdil}) to gauge
it away to an arbitrary constant:
\be
\phi^2 \ = \ -\frac{(d-1)}{4a}\ \kappa^{-2}.
\ee
In order to keep $\phi$ real one has to change the overall sign of the
action~(\ref{a2}). This leads to the usual Poincare gravity action with
gravitational constant $\kappa^{2}$. Of course, we now have to modify
the action for the conformal gauge fields, since we do not want to
keep the (higher derivative) kinetic part 
of~(\ref{fouraction}), but we do need 
the contraints~(\ref{allcon}). This is achieved by replacing the frame fields
$E_{\ul M}{}^M$ in~(\ref{sixact}) with
\be
\ce_{\ul M}{}^M \ = \ E_{\ul M}{}^K \left( 
\delta_K{}^M \ - \ \frac{U_K \Phi^M}{U_N \Phi^N} \right) \ ,
\ee 
where $\Phi_M$ is related to $\phi$ as in the previous section. 
Then at least one of the curvatures in each term of the gravitational 
part of~(\ref{comact}) carries a base space index $\ul \ominus$ and 
therefore is a Lagrange multiplier. 

With the aid of this compensator one can systematically describe any
generally relativistic system in a conformally invariant way. We find
it convenient to give the compensating scalar $f$ the scale weight
$h=-1$ by defining:
\be
\label{chvar}
\Phi_M = f^{\frac{(d-4)}{2}} f_M\ ,
\ee
$f_M$ is the new field variable and $f=U^M f_M$. By~(\ref{s}) we obtain
\be
\label{sf}
\calD f^M =-2f^M \,,\qquad \calD f=-f
\ee
and the gauge symmetry
\be
\label{varf}
\delta f^M = U^M \eta \qquad {\rm with} \qquad \calD \eta =-3\eta \ .
\ee
The action for Poincar\'e gravity in $d \geq 4$ dimensions now reads, for
example:
\bea
\label{comact}
S & = & \int_{M_{d+2}} 
\ \Bigg[ \  f^{(d-4)} \ \epsilon_{N_1 \ldots N_{d+2}} 
\ \ce^{[N_1} \wedge  \cdots \wedge  \ce^{N_{d-2}} 
 \wedge R^{N_{d-1} N_d} \wedge   R^{N_{d+1} N_{d+2}]} \nn\\
&& \phantom{\int_{M_{d+2}}}
\ + \ \frac{(d-2)}{2} |E| f^{(d-4)}
\Big( f^M U_N D_M f^N -
f^M U_M D_N f^N + \frac{d}{2} f_M f^M \Big) \Bigg]
\nn\\
&& \phantom{\int_{M_{d+2}} \ \ }
\delta(U^2)
\ \delta(\Psi) \ U^{\ul M} D_{\ul M} \Psi \,.
\eea
Invariance under~(\ref{varf}) is guaranteed for the second line since it is
a scalar action, while  the variation of the
first line is a $D$ - form $\Omega$ with $i_U \Omega=0 $ and therefore
$\Omega =0 $ by~(\ref{id}).
Note that due to the specific choice of coefficients in the
action~(\ref{scalaffact}) additional terms with derivatives of $f$ do not
appear, even though one might expect them to arise from the change of
variables~(\ref{chvar}).

We are now in a position to also generalize the action of conformal 
gravity~(\ref{sixact}) to arbitrary $d>4$:
\bea
\label{dact}
S & = & \frac{1}{8 (d-3)(d-3)!} \ \int_{M_{d+2}}
\ \epsilon_{N_1 \ldots N_{d+2}} \ E^{N_1} \wedge  \cdots \wedge  E^{N_{d-2}}
 \wedge R^{N_{d-1} N_d} \wedge R^{N_{d+1} N_{d+2}}
\nn\\ && \phantom{ - \frac{1}{(d-3)(d-3)!} \ \int_{M_{d+2}} \ }
 \ f^{(d-4)}\ \delta(U^2)
\ \delta(\Psi) \ U^{\ul M} D_{\ul M} \Psi \ .
\eea
This action can be analysed very much the same way as the action
for ordinary conformal gravity in $d=4$ in the section \ref{sec:confgrav}. 
It gives rise to the same constraints~(\ref{allcon}) and reduces to the
form \cite{tendconf}
\be
\label{a3}
S \ = \ - \frac{d-2}{4 (d-3)} \ \int_{M_{d}}|e|
\  \ \phi^{(d-4)} R(M)_{mn}{}^{pq} R(M)_{pq}{}^{mn}\ ,
\ee
where $\phi = \r f$ can be gauge fixed to a constant and
we have taken into account the constraints so that only the
Weyl part of the Rieman tensor contributes to the action. 
At the linearized level we obtain again~(\ref{calin}).
Note that for $d> 4$ the action~(\ref{a3}) is not truly
conformal (i.e. dilatation invariant) as is manifest by its
dependence on the compensator. This is in accord with the fact that
a symmetric traceless 2-index tensor does not form a 
free field representation of the conformal algebra in $d>4$.

For $D=5$ the action 
\bea
\label{con3act}
S & = & \ \int_{M_{5}} \ f
\ \epsilon_{N_1 \ldots N_{5}} \ E^{N_1}  \wedge R^{N_{2} N_3} 
\wedge R^{N_{4} N_{5}}
\ \delta(U^2)
\ \delta(\Psi) \ U^{\ul M} D_{\ul M} \Psi \ .
\eea
gives rise to the constraints~(\ref{allcon}) only. It does not describe
any dynamical gravitational field, and hence is equivalent to the
constraint part of~(\ref{comact}). In fact, it is not hard to see that
the $\Phi^M$-dependent part of $\ce^{N}$ in
(\ref{comact})  drops out in $D=5$.
Together with the second line of~(\ref{comact}) we obtain Einstein
gravity in $d=3$. 

If instead we want to describe conformal gravity in $d=3$, we must  
take the Chern-Simons action
\be
\label{5act}
S = \frac{k}{4 \pi} \int_{M_5}
 \left( \o^M{}_N  \wedge d \o^N{}_M + 
 \frac{2}{3} \o^M{}_N  \wedge \o^M{}_P  \wedge \o^P{}_M
\right) \delta(U^2)  \wedge dU^2 \ \delta(\Psi)  \wedge d \Psi \ .
\ee
It immediately reduces to the standard action by virtue of
(\ref{fourframe}):
\bea
\label{3act}
S & = & \frac{k}{4 \pi} \ \int_{M_3}
\ \left( \o^M{}_N  \wedge d \o^N{}_M \ + 
\ \frac{2}{3} \ \o^M{}_N  \wedge \o^M{}_P  \wedge \o^P{}_M
\right) \ ,
\eea
which is known \cite{hornewitten} to reproduce conformal gravity 
in $d=3$. Clearly we may write Chern-Simons actions for any semi-simple
Lie group in the same fashion. 

The Pontrjagin density is conformally invariant, and its conformal 
space version reads
\bea
\label{pont}
S & = & \frac{1}{64 \pi^2} \ \int_{M_6}
\ R^{MN}  \wedge R_{NM} 
\ \delta(U^2)  \wedge dU^2 \ \delta(\Psi)  \wedge d \Psi \ .
\eea
If one inserts the solution~(\ref{allsol}) of the constraints~(\ref{allcon}),
the result is indeed the standard Pontrjagin index (in 4 dimensions).
Again, this formula generalizes instantly to arbitrary semi-simple
Lie groups. 

The Euler density cannot, in contrast to the Pontrjagin density, be
expressed entirely in terms of conformal curvatures. It is not
conformally invariant, but of course a closed form in $M_d$.
The conformal space action therefore contains the compensator
$f^M$:
\bea
\label{euler}
S & = & \frac{1}{128 \pi^2} \ \int_{M_6} 
\ \epsilon_{N_1 \ldots N_{6}} \ U^{N_1} \ 
\widetilde f^{N_2} \wedge \widetilde R^{N_{3} N_4} 
\wedge \widetilde R^{N_{5} N_{6}} 
\ \delta(U^2)  \wedge dU^2 \ \delta(\Psi)  \wedge d \Psi
\quad \eea
with curvatures $\widetilde R^{M N} = d \widetilde \o^{MN} + 
\widetilde \o^{M}{}_K\widetilde \o^{KN}$ that arise from a modified connection
$\widetilde \o^{MN}$, and a modified compensator field
\be
\widetilde f^M \ = \ f^{-1} \left( f^M - \frac{U^M \ f^K f_K}{2f} \right) \ ,
\ee
which is invariant under~(\ref{varf}) up to trivial
terms proportional to $U^2$ and is normalized:
\be
U^M \widetilde f_M \ = \ 1 \ - \ U^2 \ \frac{f^K f_K}{2f^2} \ ,\qquad
\widetilde f^M \widetilde f_M
\ = \  U^2 \ \frac{(f^K f_K)^2}{4f^4} \ .
\ee
The modified connection is uniquely determined from the conditions
\bea
\label{lcond}
\widetilde DU^M \equiv
dU^M \ + \ \widetilde \o^{MN} U_N & = & 0
\ \quad {\rm mod} \quad
%a(y)\ \widetilde f^M \ + \ %
c^M(y) \ dU^2 \ + \ s^M(y) \ U^2
\nn\\
\widetilde D \widetilde f^M \equiv
d\widetilde f^M  \ + \ \widetilde \o^{MN}
\widetilde f_N & = & 0 \ \quad {\rm mod} \quad
%b(y)\  U^M \ + \
h^M(y)\ dU^2 \ + \ t^M(y) \ U^2
\nn\\
\eea
with arbitrary
%1-forms $a(y)$, $b(y)$,
vectors $c^M(y)$ and  $h^M(y)$
and vector-valued 1-forms $s^M(y)$ and  $t^M(y)$ :
\bea
\widetilde \o^{MN} & = & \o^{MN} 
\ - \ 2 E^{[M} \widetilde f^{N]} 
\ + \ 2 U^{[M} D\widetilde f^{N]} 
\ + \ 2 E^K \widetilde f_{K} \  U^{[M} \widetilde f^{N]} \ .
\eea
The corresponding curvature $\widetilde R^{M N}$ satisfies
as a consequence of (\ref{lcond})
\bea
\label{modcurv}
\widetilde R^{MN} \ U_N & = & 0
\ \quad {\rm mod} \quad p^{M}\ U^2
\ + \ c^{M} \ dU^2
\nn\\
\widetilde R^{MN} \ \widetilde f_N & = & 0
\ \quad {\rm mod} \quad q^{M}\ U^2
\ + \ h^{M} \ dU^2
\eea
and, when inserted into~(\ref{euler}),
may be replaced by the simpler expression
\be
\label{curvsimp}
\widetilde R^{M N} \ \rightarrow \ R^{M N} 
\ + \ E^M D\widetilde f^N \ - \ E^N D\widetilde f^M \ .
\ee
All other terms in $\widetilde R^{M N}$ cancel.  
By virtue of~(\ref{lcond}) the Euler 4-form Lagrangian 
satisfies~(\ref{gencond3})
and is therefore $\Psi$-independent,
but we note that the simpler condition~(\ref{condtop}) does not 
hold any longer.
It requires little effort to see that we reproduce indeed the usual
Euler term in $d=4$ upon gauge fixing. It is also clear that by simply 
changing the number of curvatures $R^{M N}$ and $\widetilde R^{M N}$
in~(\ref{pont}) and~(\ref{euler}) respectively, we obtain the corresponding 
topological densities in arbitrary dimensions: $d \in 2\mathbb{N}$ for
the Euler, $d \in 4\mathbb{N}$ for the Pontrjagin density. When we vary
the connections $\o^{MN}$ arbitrarily, we obtain a total derivative,
e.g. for the Euler number in $d=2$:
\be
\delta S \ = \ \frac{1}{4 \pi} \ \int_{M_4} 
\ d \left( \  \epsilon_{N_1 \ldots N_{4}} \ U^{N_1} \ 
\widetilde f^{N_2} \wedge \ \delta \o^{N_{3} N_4} \ \right)
\ \delta(U^2)  \wedge dU^2 \ \delta(\Psi)  \wedge d \Psi \ .
\quad
\ee
This equation may be integrated, and we obtain
\bea
S & = & \frac{1}{4 \pi} \ \int_{M_4} 
 \left[ \ d \left( \  \epsilon_{N_1 \ldots N_{4}} \ U^{N_1}
\ \widetilde f^{N_2} \wedge \ \o^{N_{3} N_4} \ \right)
\ + \ 2 \ \epsilon_{N_1 \ldots N_{4}} \ U^{N_1} \ 
\widetilde f^{N_2}  \ dU^{N_3} \ d\widetilde f^{N_4} \ \right]
\nn\\ &&
\phantom{ \frac{1}{4 \pi} \ \int_{M_4} }
\ \delta(U^2)  \wedge dU^2 \ \delta(\Psi)  \wedge d \Psi \ .
\quad
\eea
We recognize in the first term the straightforward extension of
the Chern-Simons density to conformal space.   
The second term is unfamiliar, but is readily understood
if one observes that due to~(\ref{curvsimp}) there are $\o^{MN}$ - independent
terms in~(\ref{euler}). These terms are similar to Hopf invariants, and appear
also in the (A)dS gauge theory formulation of gravity \cite{preitvas}. 

\section{Vector- and p-form Fields}\label{sec:vector}

We describe Yang-Mills gauge fields in $d$ dimensions by means
of a $(d+2)$-dimensional vector potential
$A_{\ul N}$ with field strength
\be
F_{{\ul N} {\ul M}} = \partial_{{\ul N}} A_{{\ul M}}
-\partial_{{\ul M}} A_{{\ul N}} +[ A_{\ul N} , A_{\ul M}]\ ,
\ee
where $A_{\ul N}$ and $F_{{\ul N} {\ul M}}$ take values in some
semi-simple Lie algebra $g$.
We impose the standard transversality condition~(\ref{transf})
and choose the action in the form
\be
\label{vectact}
 S \ = \ -\frac{1}{2g^2}\int_{M_{d+2}} |E|\ f^{(d-4)}
\ G^{{\ul N} {\ul M}} \ G^{{\ul R} {\ul P}}
\ tr \left( F_{{\ul N} {\ul R}} F_{{\ul M} {\ul P}}\right) 
\ \delta(U^2) \ \delta(\Psi) \ U^{{\ul M}} D_{{\ul M}} \Psi\, ,
\ee
where $G^{{\ul N} {\ul M}}$ is the  $(d+2)$-dimensional
metric tensor~(\ref{sixmetr}).
This action as well as the constraint~(\ref{transf}) is obviously invariant 
under the  ordinary Yang-Mills gauge transformations
\be
\label{ga}
\delta A_{{\ul N}} = D_{{\ul N}} \L \ ,
\ee
where $\L(Y)$ is an arbitrary parameter with scale weight $h_\L = 0$ 
taking values in $g$. We may choose as a special case
\be
\label{gaspec1}
\L = \half U^2 \S \quad \Longrightarrow \quad
\delta A_{\ul M} \Big|_{U^2=0} \ = \ U_{\ul M} \S \Big|_{U^2=0} \ ,
\ee
with $h_\S = -2$. This symmetry is analogous to that of the scalar
case~(\ref{gscal}) and ensures that the component of the gauge vector
proportional to $U_{\ul M}$ does not appear in the action. 
Another special case is 
\be
\label{gaspec2}
\L = \half U^2 \ U^{\ul M} S_{\ul M} \quad \Longrightarrow \quad
\delta A_{\ul M} \ = \ U^2 \ S_{\ul M} \ + \ 2 U_{\ul M} \ U^{\ul N} S_{\ul N} 
\ + \ U^2 \ U^{\ul N} D_{\ul M} S_{\ul N}\ ,
\ee
which is the analog of the symmetry~(\ref{restrict}). 

One may wonder about the straightforward generalization of the gauge field
strength to conformal space, since following the first quantization approach 
of \cite{siegel} one would expect a field strength
${\cal H}_{\ul K \ul L \ul M}$ that satisfies 
\be
\label{ufconstr}
U^{\ul K}{\cal H}_{\ul K \ul L \ul M}=0 \qquad {\rm and} \qquad    
U_{[\ul J}{\cal H}_{\ul K \ul L \ul M]}=0 \ . 
\ee
Only for such field strengths
can one define a selfduality condition in 
$D=6$ conformal space, which translates
to selfdual field strengths in physical spacetime, and they are
precisely the irreducible free field representations of the conformal  
algebra, and therefore we wish to have them at our disposal. 
The solution to the contraints~(\ref{ufconstr}) is 
\be
\label{ufsolv}
{\cal H}_{\ul K \ul L \ul M} = U_{[\ul K}F_{\ul L \ul M]} \ , 
\ee 
but 
this quantity is not useful for constructing an action. Instead,
we will use a field ${\cal F}_{\ul K \ul L \ul M}$ that is only
constrained by the scaling condition 
$\calD {\cal F}_{\ul K \ul L \ul M}=0$
in the action
\bea
\label{vectact2}
S & = & \frac{1}{2g^2}\int_{M_{d+2}} |E|\ f^{(d-4)}
\ tr \left( 2 {\cal F}^{\ul K \ul L \ul M} U_{\ul K} F_{\ul L \ul M}  
\ + \ U_{\ul K} {\cal F}^{\ul K \ul L \ul M} 
\ U_{\ul J}{\cal F}^{\ul J}{}_{\ul L \ul M} \right)
\nn\\ && \phantom{ -\frac{1}{4g^2}\int_{M_{d+2}} } 
\ \delta(U^2) \ \delta(\Psi) \ U^{\ul M} \partial_{\ul M} \Psi
\ , \eea
in which again the differential operator $U_{[\ul K} D_{\ul L]}$ appears.
The equations of motion that follow from~(\ref{vectact2}) are on the physical
hypersurface precisely those one would expect for~(\ref{ufsolv}).  

The form of the frame field~(\ref{infr}) and the transversality
condition immediately imply (for either action) that
\be
\label{vectact4}
 S \ =\ -\frac{1}{4g^2}\int_{M_d} |e|\ \phi^{(d-4)} 
\ g^{{\ul n} {\ul m}}\  g^{{\ul r} {\ul s}} 
\ tr(F_{{\ul n} {\ul r}} F_{{\ul m} {\ul s}}) \ ,
\ee
where $g^{{\ul n} {\ul m}}$ is the metric tensor constructed from
$e_n{}^{\ul n}$. After the compensator $\phi$ is fixed to some
(dimensionful) constant one arrives at the standard Yang-Mills action
in $d$ dimensions. The case of $d=4$ is conformal, since then the
action becomes independent of the compensator.

Another important ingredient in Yang-Mills actions are topological
terms. By their very nature they are conformally invariant. In conformal
space we may write them as
\bea
\label{gatop} 
S & = & \int_{M_{d+2}} \ E^N U_N  
\wedge tr \left( F  \wedge \cdots \wedge F \right) 
\ \delta(U^2) \ \delta(\Psi) \wedge d \Psi
\nn\\ & = &
\frac{1}{2} \ \int_{M_{d+2}} \ tr \left( F  \wedge \cdots \wedge F \right) 
\ \delta(U^2)  \wedge d U^2 \ \delta(\Psi)  \wedge d \Psi \  ,
\eea
where $tr$ may be replaced by any invariant tensor of the group in question.
We obviously obtain the standard topological term in $d$ dimensions, 
i.e. the $\theta$-term in $d=4$.

Gauge interactions for conformal matter are desribed by simply
gauge covariantizing all derivatives:
\be
\partial_{\ul N} \F \ \rightarrow  
\ \nabla_{\ul N} \F \ \equiv \ \partial_{\ul N} \F + A_{\ul N}(\F) \ ,
\ee
with $A_{\ul N}$ taking values in the appropriate representation of the 
Lie algebra $g$.

In the same fashion in which we just discussed vector fields one
may also describe p-form gauge fields. We select the action
\bea
\label{formpact}
 S & = & - \frac{1}{(p+1)!} \int_{M_{d+2}} \ |E|
\ f^{d-2(p+1)}
\ G^{{\ul M_1} {\ul N_1}} \cdots G^{{\ul M_{p+1}} {\ul N_{p+1}}}
\ H_{{\ul M_1} \cdots {\ul M_{p+1}}}\ H_{{\ul N_1}\cdots {\ul  N_{p+1}}} 
\nn\\ && \phantom{ - \frac{1}{(p+1)!} \int_{M_{d+2}} \ }
\ \delta(U^2) \ \delta(\Psi) \ U^{{\ul M}} D_{{\ul M}} \Psi \ ,
\eea
with totally antisymmetric (p+1)-form field strength
\be
H_{ \ul M_1 \cdots \ul M_{p+1} } \ =
 \ \del_{\ul M_1} A_{ {\ul M_2} \cdots {\ul M_{p+1}} } \ \pm 
\ (\ p \ {\rm more \ terms}\ )
\ee
obeying $U^{\ul M_1} H_{ \ul M_1 \cdots \ul M_{p+1} } = 0$ 
or in form notation $H = (p+1)\ dA$, $i_U H = 0$.
This yields in d dimensions
\be
\label{formactd}
 S \ =\ -\frac{1}{2 \ (p+1)!}\int_{M_d} |e|\ \f^{d-2(p+1)}
 \ g^{{\ul m_1} \cdots {\ul n_1}} \  g^{{\ul m_{p+1}} {\ul n_{p+1}}}
 \ H_{{\ul m_1} \cdots {\ul m_{p+1}}}
H_{{\ul n_1} \cdots {\ul n_{p+1}}} \ ,
\ee
and we remark that as expected, for $d/2= p+1$ the compensator fields
drop out,
signalling true conformal symmetry.

\section{Fermions in d=(1,3)}\label{sec:fermion}

Spinor fields $\jb_a, \psi^a$ transform
under $Spin(2,4)=SU(2,2)$. We use the following
conventions: $\psi^a = (\psi^\a, \ub^{\da})$,
$\jb_a = ( \u_\a , \jb_{\da} )$,
$U^{ab}=U_M \S^M{}^{[ab]}$,
$U_{ab}=U_M \S^M{}_{[ab]}$,
\be
\S^M{}^{[ab]} \ = \ \left(
\begin{array}{cc}
\sqrt{2} \et^{\a\b} \de^M_{\oplus} & \s^{m \a \db} \\
-  \s^{m \da \b} & \sqrt{2} \et^{\da\db} \de^M_{\ominus}
\end{array}
\right)
\quad
\S^M{}_{[ab]} \ = \ \left(
\begin{array}{cc}
\sqrt{2} \e_{\a\b} \de^M_{\ominus} & \s^{m}_{ \a \db} \\
-  \s^{m}_{\da \b} & \sqrt{2} \e_{\da\db} \de^M_{\oplus}
\end{array}
\right)\,,
\ee
where $\s^{m}_{ \a \db}$ are $SL(2,C)$- sigma-matrices with
$\s^{m \a}{}_{\db} \sb^{n \db}{}_{\g}= \de^\a_\g \h^{mn}
+ \s^{[mn]}{}^{\a}{}_{\g}$ and $\et^{12}=-\epsilon_{12}=1$.
Then
\be
\S^{M}{}^{[ab]} \S^{N}{}_{[bc]} \ = \ \eta^{MN} \de^a{}_c \ +
 \ \S^{[MN]}{}^a{}_c \ .
\ee
The covariant derivative for spinors reads
\be
D_{\ul M} \psi^a \ = \ \del_{\ul M}  \psi^a \ + \ \frac{1}{4}
\o_{\ul M}{}^{MN} \S_{[MN]}{}^a{}_b \psi^b
\ee
and if $\psi^a$ carries a representation of an additional
internal Yang-Mills gauge group, we denote it by
\be
\nabla_{\ul M} \psi^a \ = \ D_{\ul M}  \psi^a \ + \ A_{\ul M} (\psi^a) \ ,
\ee
where $A_{\ul M}$ is a Lie-algebra valued vector gauge field, and the
scaling operator $\calD$ is in that case defined to be Yang-Mills
gauge covariant. We identify the physical components of the spinor
fields $\jb_a, \psi^a$ with those invariant under the transformation
\be
\label{g1f}
\de \jb_a \ = \ U_{ab} \U^b \quad , \ \de \psi^a \ = \ U^{ab} \Ub_b \ ,
\label{spinsymm}
\ee
i.e. with the spinors
\be
\chi^a \ = \ U^{ab} \jb_b \quad , \ \cb_a \ = \ U_{ab} \psi^b\ ,
\ee
(on the  hypersurface $U^2=0$) and impose the scaling property
\bea
\calD \jb_a \ = \ h \jb_a & \quad \calD \psi^a \ = \ h \psi^a\ ,\nn \\
\calD \U^a \ = \ (h-1) \U^a & \quad \calD \Ub_a \ = \ (h-1) \Ub_a \,.
\eea
With $\nabla_{ab} \psi^b = \S^M{}_{[ab]} E_M{}^{\ul L} \nabla_{\ul L} \psi^b$
the action
\be
S = \int_{M_6} |E| \ \frac{i}{\sqrt{2}} 
\ [ \ \cb_a \nabla^{ab} \jb_b - \chi^a \nabla_{ab} \psi^b \ ] 
\ \delta(U^2) \ \delta(\Psi) \  U^{\ul M} D_{\ul M} \Psi
\label{fermiaction}
\ee
is invariant for $h=-2$, by virtue of
\be
U_{ab} \nabla^{bc} U_{cd} \U^d
\ = \ -U^2  \nabla_{ab} \U^b + (6+2(h-1))\ U_{ab} \U^b  \
\ee
under  the symmetry (\ref{g1f}) analogous to
(\ref{gscal}) for scalars, as well as under the symmetry
\be
\label{g2f}
\delta \psi^a \ = \ U^2 \ \Xi^a\,,\qquad
\delta \bar{\psi}_a \ = \ U^2 \ \bar{\Xi}_a \,
\ee
analogous to (\ref{restrict}).

One may wonder about the uniqueness of~(\ref{fermiaction}). After all, we are
not allowed to partially integrate in the Lagrangian density because
it contains nontrivial delta-functions, and therefore
terms like $\jb_a \nabla^{ab} \cb_b$ are to be considered independent.
Besides, also $\jb_a \psi^a$ satisfies the correct scaling condition.
The only combination of those terms that is invariant under~(\ref{spinsymm}) 
turns out to be
\be
\jb_a \nabla^{ab} \cb_b - (6+2h)\ \jb_a \psi^a \ = 
\ \chi^a \nabla_{ab} \psi^b \ .
\ee
and therefore is already included in~(\ref{fermiaction}). Hence
the action (\ref{fermiaction}) is essentially unique.
Yukawa couplings to scalars are given (for one real boson) by
\be
S^{int} = \int_{M_6} |E| \ \frac{1}{\sqrt{2}}
\ U^M \Phi_M \ [\  \jb_a U^{ab} \jb_b +  \psi^a U_{ab} \psi^b \ ]
 \ \delta(U^2) \ \delta(\Psi) \  U^{\ul M} D_{\ul M} \Psi \ .
\label{yukawas}
\ee
This action is invariant under the symmetries
 (\ref{g1f}), (\ref{g2f}), (\ref{gscal}) and
 (\ref{restrict}).

We should now make sure that~(\ref{fermiaction}) does yield
the usual Lagrangian for fermions in d=(1,3).
To that end we use $U_{ab} | = \r\ \S^{\ominus}{}_{ab}$ as well as
the inverse framefield~(\ref{infr}) and observe:
\bea
\chi^a \nabla_{ab} \psi^b \Big| & = &
   - \ \jb_a U^{ab} \nabla_{bc} \psi^c  \Big|  \nn \\
& = &   - \ \jb_a \ \S^{\ominus}{}^{ab}
 \   \Bigg[ \S^{\oplus}{}_{bc} \nabla_{\ul \oplus} \psi^c
         + \S^{\ominus}{}_{bc}  \nabla_{\ul \ominus} \psi^c
\nn \\
&&
\phantom{  - \psi^a \ \S^{\ominus}{}_{ab} \ \Bigg[\Bigg] }
         + \S^{k}{}_{bc} \left( \nabla_{k} \psi^c
         - \left\{ \del_k \ln \r + \o_k{}^{\oplus\ominus} \right\}
                  \nabla_{\ul \oplus} \psi^c   \right)
   \Bigg]  \ .
\eea
The field $\nabla_{\ul \ominus} \psi^c$ is projected out,
$\nabla_{\ul \oplus} \psi^a = -2 \psi^a$ and
\be
\nabla_{k} \psi^\a
 \ = \ e_k{}^{\ul k}\left (
\del_{\ul k} \psi^\a + A_{\ul k} (\psi^\a) + \frac{1}{4}
\o_{\ul k}{}^{mn}
\s_{mn}{}^\a{}_\b  \psi^\b
- \half \o_{\ul k}{}^{\oplus\ominus}  \psi^\a \right ) -
\divrt \s_k{}^\a{}_{\db} \ub^{\db} \ ,
\ee
so that finally
\bea
\chi^a \nabla_{ab} \psi^b \Big| & = &
\sqrt{2} \ \jb_\da\ \sb^k{}^{\da}{}_\b \ \left( D^{\cal L}_{k} + A_k +
                    \frac{3}{2}  \o_k{}^{\oplus\ominus}
                    + 2 \del_k \ln \r \right) \psi^\b \nn \\
\cb_a \nabla^{ab} \jb_b \Big| & = &
- \sqrt{2}\ \psi^\a\ \s^k{}_{\a}{}^{\db}\ \left( D^{\cal L}_{k} + A_k +
                    \frac{3}{2}  \o_k{}^{\oplus\ominus}
                    + 2 \del_k \ln \r \right) \jb_\db \ .
\eea
Remarkably, the scale weight $h=-2$ and the eigenvalue of the
dilatation generator in tangent space assemble to yield the proper
conformal weight $3/2$ for fermions. Again we rescale $\psi_\a$,
and like in the scalar case $\o_k{}^{\oplus\ominus}$
does not appear in the four-dimensional action
\be
S = \int_{M_4} |e| \ \frac{i}{2}
 \ [ \  \psi^\a \s^k_{\a \db}\ D^{\cal L}_{k} \jb^\db
-  D^{\cal L}_{k} \psi^\a \s^k_{\a \db}\ \jb^\db \ ] \ .
\label{fourfermiaction}
\ee
Using the same gauge fixing procedure, the Yukawa interaction is brought
to the form
\be
S^{int} = \int_{M_4} |e| \ \half \ \phi
 \ [ \  \psi^\a  \psi_\a +  \jb_\da  \jb^\da \ ] \  .
\label{fouryukawa}
\ee

\section{Gravitinos in d=(1,3)}\label{sec:gravitino}

We treat gravitino fields as fermionic gauge fields,
with field strengths
\bea
R_{\ul M \ul N}{}^a & = & D_{\ul M} \j_{\ul N}{}^a -
                          D_{\ul N} \j_{\ul M}{}^a \\
\Rb_{\ul M \ul N}{}_a & = & D_{\ul M} \jb_{\ul N}{}_a -
                          D_{\ul N} \jb_{\ul M}{}_a \quad ,
\eea
which are chosen transversal:
\be
U^{\ul M} R_{\ul M \ul N}{}^a \ = \ 0 \ = \ U^{\ul M} \Rb_{\ul M \ul N}{}_a
\ .
\ee
We note the decomposition
\be
R^a \ = \ \left(
\begin{array}[c]{c}
R^\a(Q) \\
\Rb^\da(S)
\end{array} \right)
\ = \ \left(
\begin{array}[c]{c}
 2 (D \j)^\a \\
 2 (D \fb)^\da
\end{array} \right)
\quad ; \quad
\Rb_a \ = \ \left(
\begin{array}[c]{c}
R_\a(S) \\
\Rb_\da(Q)
\end{array} \right)
\ = \ \left(
\begin{array}[c]{c}
 2 (D \f)_\a \\
 2 (D \jb)_\da
\end{array} \right)
\ee
with
\be
\begin{array}[c]{ccccl}
\half R(Q)^\a & = & d\j^\a +
     \frac{1}{4} \o^{mn} \s_{mn}{}^\a{}_\b \j^\b +
     \half b \j^\a - \frac{1}{\sqrt{2}} e^m \s_{m}{}^{\a}{}_\db \fb^\db \\
\half R(S)_\a & = & d\f_\a +
     \frac{1}{4} \o^{mn} \s_{mn}{}_\a{}_\b \f^\b -
     \half b \f_\a + \frac{1}{\sqrt{2}} f^m \s_{m}{}_{\a}{}_\db \fb^\db \ .\\
\end{array}
\ee

The action reads
\be
\label{gravitinoact}
S \ = \int_{M_{6}} |E|
\left\{\
8 \ i \ R_{MN}{}^a \ \Rb^{MN}{}_a
 \ + \ i \ \S^N_{ab}\ R_{NM}{}^b \ \S_L^{ac}\ \Rb^{LM}_c
\ \right\}
\delta(U^2)
\ \delta(\Psi) \ U^{{\ul M}} D_{{\ul M}} \Psi\ .
\ee
Observing $R_{\oplus M}{}^a \Big| = 0 = \Rb_{\oplus M}{}_a \Big|$,
we obtain for the constraint part
\bea
S_{\rm constraint} & = & \ \int_{M_{6}} i \ |E| \ \delta(U^2)
\ \delta(\Psi) \ U^{{\ul M}} D_{{\ul M}} \Psi
\nn\\
& & \phantom{ \ \int_{M_{6}}}
\Big\{
R_{nm}{}^\b(Q) \ \s^n{}_{\b\da}
\ \left( \ -\sqrt{2} \Rb_{\ominus}{}^{m\da}(Q)
\ + \ \sb_{p}{}^{\da\g} R^{pm}{}_\g(S) \ \right) \nn\\
& & \phantom{ \ \int_{M_{6}}}
 + \ \left( \ \sqrt{2} R_{\ominus m \a}(Q)
\ + \ \s^{p}{}_{\a\db} \Rb_{pm}^\db(S) \ \right)
\ \s_n{}^{\a\dg}\ \Rb^{nm}{}_\dg(Q)
\ \Big\} \ .
\eea
The fields $R_{\ominus m}{}^{\a}(Q) =
\del_{\ul \ominus} \j^\a + \cdots $ and 
$\Rb_{\ominus m}{}^{ \da}(Q) = \del_{\ul
\ominus} \jb^\da + \cdots$ may be regarded as independent fields that play
the role of Lagrange multipliers for the standard constraints
\be
\s^m{}_{\a\db} \ \Rb_{mn}{}^\db(Q)  \ = \ 0 \quad ; \quad
\sb^m{}_{\da\b} \ R_{mn}{}^\b(Q)  \ = \ 0  \ ,
\ee
which imply in particular
\bea
\Rb_{mn}{}^\db(Q)  & = & -\frac{i}{2} \ \e_{mn}{}^{pq}  \Rb_{pq}{}^\db(Q)
\nn\\
R_{mn}{}^\b(Q)  & = & \frac{i}{2} \ \e_{mn}{}^{pq}  R_{pq}{}^\b(Q) \ .
\eea
The (kinetic part of the) action now takes the well-known form
\be
S \ = \ 2 \ \int_{M_{4}} 
\ \e^{mnpq}\ \left( \ R_{mn}{}^\a(Q)\ R_{pq}{}_\a(S)   
\ - \ \Rb_{mn}{}^\da(Q)\ \Rb_{pq}{}_\da(S) \ \right)  .
\ee
We may couple the gravitinos in the standard way to a $U(1)$ gauge symmetry,
and obtain the action of conformal supergravity in conformal space. 

\section{Conclusions and Outlook}\label{sec:conclusions}

We have presented the theory of conformal gravity as a gauge theory
of the conformal group in local conformal space. In order to define 
physical spacetime as a hypersurface of codimension 2 in this conformal space,
we introduced a field $U^M(y)$ which allowed us to define the
local cone $U^M(y)U_M(y)=0$ as well as the projectivity condition 
$U^{\ul M} \partial_{\ul M}=h$ in a gauge- and reparametrization-invariant
way. One might
interprete this field $U^M(y)$ as a compensator for the
conformal group, but as we have shown this field remains invariant
under global (vacuum) conformal symmetries.
It also may be viewed as the generalization of
the coordinate $y^{\ul M}$ for nontrivial gravitational fields.
This has profound consequences: in the first-quantized action
that describes conformal particles \cite{hppt} we simply replace $y^{\ul M}$ by
$U^M(y)$ in order to couple to a nontrivial background:
\be
S  =  \int d\t \Big[ 
\ \half D_\t U^M  D_\t U_M 
\ + \half \lambda U^2\ \Big] \ 
\ee
with $D_\t U^M = \partial_\t U^{M} +  
\partial_\t y^{\ul M} \o_{\ul M}{}^M{}_N U^N$.  
In order to make contact with the standard formulation of particle
quantum mechanics, one has to use the key property 
$D_{\ul M} U^N = E_{\ul M}{}^N$. The importance of this soldering
form was already recognized by Stelle and West in their treatment of
AdS gravity \cite{stellewest}. Here it is used for conformal space,
and we believe that it will be useful in a much wider context: one
may generalize the base space to a superspace, for example, one may 
generalize the fibre to some supergroup, as we have done for conformal
supergravity \cite{cpandmv}, or one may generalize the particle 
worldline to a string worldsheet, or to a p-brane worldvolume:
\bea
S &  = & \int d^{p+1}\xi 
\sqrt{g} \left( \half g^{\a\b}  D_\a U^M  D_\b U_M + 
\l U^2 + \half (p-1) \right) \ .
\eea

It is remarkable how naturally the constraints of conformal 
(super)gravity appear in the framework of conformal space. 
They are enforced by fields that have their origin in one of
the extra dimensions: these Lagrange multipliers are differential  
forms which are partially transverse to the physical hypersurface.

We conclude, therefore, that the framework of local conformal space
is the correct setting for the description of theories with local
conformal symmetry, which may be spontaneously broken e.g. by 
extra compensators.

\section*{Acknowledgments}

We are grateful to a number of people and institutes who helped
in our collaboration, notably 
L. Brink at Chalmers Tekniska H\"ogskola and G\"oteborg Universitet, 
B. de Wit at Utrecht University, 
G. Ferretti at SISSA, 
A. Gurevich at Lebedev Institute, Moscow, 
E. Ivanov at the JINR, Dubna,
H. Nicolai at the Albert-Einstein-Institute in Potsdam and
D. L\"ust at Humboldt-Universit\"at zu Berlin.
C.P. thanks P. van Nieuwenhuizen for valuable lessons in 
conformal gravity and supergravity. \\
This work was supported in part by
INTAS grants CT93-0023,
 96-538, RBRF grant 96-02-17314,
Swedish Natural Science Research Council grant F-FU 08115-342
and DFG project 436 RUS 113.

\appendix
\setcounter{section}{1}
\setcounter{equation}{0}

\section*{Notation and Conventions}\label{sec:conventions}

\subsection*{Symmetrization}

For all index types
symmetrizations and antisymmetrizations are projectors, e.g. 
\bea
T^{(mn)} & = & \half\left(T^{mn} + T^{nm}\right) \nn\\
T^{[mn]} & = & \half\left(T^{mn} - T^{nm}\right) \quad .
\eea
Gamma- or Sigma-matrices are antisymmetrized in the same fashion, e.g.
\be
\G^{mn} \ = \ \G^{[m}\G^{n]} \ = \ \half \left( \G^m \G^n - \G^n \G^m \right)
\ee

\subsection*{Conformal Space and $SO(2,d)$}

\be
\Psi_{\ul M} = E_{\ul M}{}^M \Psi_{M} \quad ; \quad  
\Psi^{\ul M} =  \Psi^{M} E_{M}{}^{\ul M} \quad ; \quad  
\Psi^{M} \Psi_{M} = \Psi^{\ul M} \Psi_{\ul M}
\ee
$M$ are $SO(2,d)$ vector indices indices, with
metric $(- + + + \cdots + -)$ with indices $M\in {0,1,2,3,\cdots,d,d+1}$,
note $\eta_{dd} = 1 = \eta^{dd}$, 
$\eta_{(d+1)(d+1)} = -1 = \eta^{(d+1)(d+1)}$.
We define
\bea
A^{\oplus} \ = \ \frac{1}{\sqrt{2}}(A^d + A^{d+1}) 
\ = \ A_{\ominus} \ = \ \frac{1}{\sqrt{2}}(A_d - A_{d+1}) \nn\\
A_{\oplus} \ = \ \frac{1}{\sqrt{2}}(A_d + A_{d+1}) 
\ = \ A^{\ominus} \ = \ \frac{1}{\sqrt{2}}(A^d - A^{d+1}) \nn\\
\eea
and then 
\be
A^N B_N \ = \ A^{\oplus} B^{\ominus} +  A^{\ominus} B^{\oplus} + A^n B_n \ .
\ee

$\ul M$ are $d+2$-dimensional world indices:
\be
y^{\ul M} \ = \ ( y^{\ul \oplus}, y^{\ul \ominus}, x^{\ul m})\,.
\ee
For simplicity, we consider four dimensions in the following, in which case
$m$ are $SO(1,3)$ indices and $\ul m$ are 4-d world indices. Our 
integration conventions are:
\be
\int  dy^{\ul \oplus}\ dy^{\ul \ominus}\ dx^{\ul m} \ \delta(y^{\ul \oplus})
\ = \ \int dy^{\ul \ominus}\ dx^{\ul m}\,.
\ee
In conformal space we define the completely antisymmetric tensor as follows:
\bea
\e_{540123} \ = \ 1 \quad &,& \quad\e^{540123} \ = \ 1 \nn\\
\e_{\oplus \ominus 0123} \ = \ 1 \quad &,& \quad\e^{\oplus \ominus 0123} 
\ = \ 1 \ ,
\eea
the Minkowski counterpart reads:
\be
\e_{0123} \ = \ -1 \quad, \quad\e^{0123} \ = \ 1 \ .
\ee

\section*{Scalar Actions}\label{scal}

We start with an action of the form
\bea
\label{Ascalansatz}
S & = & \int_{M_{d+2}} |E|
\left[ a \varphi D_M\Phi^M +
       b \Phi^M D_M \varphi +
       c \Phi_M \Phi^M +
       f D_M\varphi D^M\varphi +
       g \varphi D_M D^M \varphi
\right]
\nn\\
&& \phantom{ \int_{M_{d+2}} |E| }
\ \delta(U^2)
\ \delta(\Psi) \  U^{\ul M} \partial_{\ul M} \Psi \,,
\eea
where $a,b,c,f$ and $g$ are arbitrary real parameters.

The action is invariant under the transformation~(\ref{gscal}) provided that
\be
\label{scalginv}
c=-\frac{(d+2)}{4}a +\frac{(d-6)}{4} b + (d-2) f\ .
\ee
The invariance under (\ref{restrict}) requires
\be
\label{tun}
a+b= (d-2)f\ ,
\ee
and then $c= \frac{1}{4} (d-2)(b-a)$.
In $d \not= 2$ the ``natural'' action where only $f \not= 0$
is nondynamical, and in fact completely trivial,
since the condition~(\ref{tun})
is not satisfied.

After
imposing~(\ref{scalginv}) and~(\ref{tun}) we are left with:
\bea
\label{Ascalact}
 S & = & \int_{M_{d+2}} |E|
\Bigg[
\ a \ \Phi^M U_M D_N \Phi^N
\ + \ \frac{2a + db}{d-2}\ \Phi^M U_N D_M \Phi^N
\nn\\ && \phantom{ \int_{M_{}}}
\ + \ \frac{d^2 (b-a) + 4 d a}{4(d-2)}\ \Phi_M \Phi^M
\ + \ \frac{a+b}{d-2}\ U^K D_M \Phi_K \ U^N D^M \Phi_N
\nn\\ && \phantom{ \int_{M_{}}}
\ + \ g \ \Phi^M U_M \left( 2  D_N \Phi^N + U^K D^N D_N \Phi_K \right)
\Bigg]
\ \delta(U^2)
\ \delta(\Psi) \ U^{\ul M} D_{\ul M} \Psi \ .
\eea
For $a=0$, $g=0$ the action is nondynamical
 for any choice of coefficients $b,c$, since then~(\ref{Ascalact})
takes the form
\bea
\label{nondyn}
 S & = & \int_{M_{d+2}} |E| \ \frac{b (d-2)}{4}
\ \Big( \Phi_M \ + \ \frac{2}{d-2} D_M \varphi \Big)^2
\ \delta(U^2)
\ \delta(\Psi) \ U^{\ul M} D_{\ul M} \Psi \ ,
\eea
and we obtain the extra symmetry
\be
\de \Phi^M \ = \ - \frac{2}{d-2} D_M \L  \quad , \quad
\de \varphi \ = \ \L \ .
\ee
We may now gauge fix $\varphi$ to zero and then it is obvious
that~(\ref{nondyn}) does
not describe dynamical degrees of freedom.

By a field redefinition
\be
\label{fieldredef}
\Phi_M \longrightarrow \Phi_M + \a D_M \varphi
\ee
we change $f \longrightarrow f - \a a$
(we use here~(\ref{scalginv}) and~(\ref{tun})),
as well as
\bea
a & \longrightarrow & a - \frac{d-2}{2}\a a \nn\\
b & \longrightarrow & b - \frac{d-2}{2}\a a \nn\\
g & \longrightarrow & g \left( 1 -  \frac{d-2}{2}\a \right)^2
+ \left( 1 -  \frac{d-2}{2}\a \right) \a a\ ,
\eea
and hence we may set $f=0$ unless $a=0$. Let us note that the
field redefinition describes a shift of d-dimensional
auxiliary fields by a derivative of the dynamical field $\varphi$
and therefore it does not affect a structure of the
physical phase space.

Up to a total derivative and a field redefinition~(\ref{fieldredef})
the affine action~(\ref{scalaffact}) is in fact equivalent to the general
case~(\ref{Ascalact}). Naively the delta-functions in~(\ref{Ascalact}) would
seem to prohibit us from introducing the concept of partial integration,
but consider the following action of topological type:
\bea
\label{delsca1}
\Delta S & = & \b \frac{ (-)^{d-1}}{(d-1)!}\int_{M_{d+2}}
\delta (U^2 )\ dU^2 \ \delta (\Psi) \wedge d \Psi \
\nn \\ && \phantom{\frac{(-)^{d-1}}{(d-1)!}\int_{M_{d+2}} }
\wedge d \ \epsilon_{N_1 \ldots  N_{d+2}}
\wedge E^{N_1} \wedge \ldots E^{N_{d-1}}
\ U^{N_d} \ D^{N_{d+1}}\varphi \ \Phi^{N_{d+2}} \ .
\eea
It is manifestly a total derivative and satisfies the
symmetry requirements~(\ref{gscal}) and~(\ref{restrict})
for a scalar action.
After some computation we obtain, using~(\ref{allconinv}),
\bea
\label{delsca2}
\Delta S & = & 2\b  \int_{M_{d+2}} |E|
\Bigg[
\ \varphi \ D_M  D^M \varphi \ + \   D_M \varphi \ D^M \varphi
\nn\\ && \phantom{\int}
\ + \ \left( \frac{d}{2} -1 \right)
\Big( \ \Phi^M D_M \varphi
\ + \ \varphi D_M \Phi^M \ \Big)
\Bigg]
\ \delta(U^2)
\ \delta(\Psi) \ U^{\ul M} D_{\ul M} \Psi \ ,
\eea
and with an appropriate choice of coefficients $\a,\b$ in~(\ref{fieldredef})
and~(\ref{delsca1}) we may set $f=g=0$, which implies the
form~(\ref{scalaffact}) of the scalar action. Alternatively, we may
choose $a=f=0$ and work with a simple $\varphi \Box \varphi$ - type action.

We will now show directly
that~(\ref{Ascalact}) describes a conformally coupled
scalar field in $d$ dimensions. Imposing the gauge conditions
(\ref{p+})
and (\ref{o+}),
we reduce~(\ref{Ascalact}) to the form:
\bea
 S & = & \frac{1}{2} \int_{M_d} \r^d |e|
\Bigg[
\ a \ \r \Phi_\oplus  D_n \Phi^n
\ + \  \frac{2a + db}{d-2}\ \Phi^n \r  D_n \Phi_\oplus
\nn\\ && \phantom{\frac{1}{2} \int_{M_d} \r^d |e| \ }
\ + \  \frac{d^2 (b-a) + 4 d a}{4(d-2)}\ \Phi_n \Phi^n
\ + \  \frac{a+b}{d-2} \ \r^2\ D_n \Phi_\oplus D^n \Phi_\oplus
\nn\\ && \phantom{\frac{1}{2} \int_{M_d} \r^d |e|}
\ + \ g \ \Phi_\oplus \left( -d \r D_\ominus \Phi_\oplus + 2 \r D_n \Phi^n
+ \r^2 D^n D_n \Phi_\oplus \right)
\Bigg]\,.
\eea
Making use of  (\ref{re}) as well as
\bea
\r^2 D^n D_n \Phi_{\oplus} & = &  \r^{-d/2}
\Bigg\{
  \frac{1}{e} \partial_{\ul n}
    \left( e e_n{}^{\ul n}
        \left[ \partial^n \phi - \frac{1}{2}(d-2) b^n - \phi^n
        \right]
    \right)
\nn\\ && \phantom{ \r^{-d/2} \Bigg\{ }
    - \frac{d}{2} \o_n{}^{n \oplus} \phi
    + \frac{1}{2}(d-2)b^n \left( \partial_n \phi -
                              \frac{1}{2}(d-2) b_n - \phi_n \right)
\Bigg\}
\nn\\ &&
+ \ d \r D_\ominus \Phi_\oplus - \r D^n \Phi_n
\eea
after redefinition (\ref{resc}) the action reduces to the form
\bea
\label{Aa1}
S & = &  \frac{1}{2} \int_{M_d} |e|
\ \Bigg\{ \ \frac{a+b}{d-2} \partial_n \phi \ \partial^n \phi
\ - \ (a+b) \ b^n \ \phi \ \partial_n \phi
\ + \ \frac{1}{4} (d-2)(a+b) \ b^n b_n \ \phi^2
\nn\\ && \phantom{\frac{1}{2} \int_{M_d} |e|}
\ + \ a \ \o_n{}^{n \oplus} \ \phi^2
\ + \ (b-a) \ \phi^n
\ \left[
\partial_n \phi - \frac{1}{2}(d-2) b_n \phi + \frac{1}{4} (d-2) \phi_n
\right]
\nn\\ && \phantom{\frac{1}{2} \int_{M_d} |e| \ \Bigg\{ \ }
+\ g \phi \ \Bigg(  \frac{1}{e} \partial_{\ul m} \left[ e e_n{}^{\ul n}
(\partial^n \phi - \frac{1}{2}(d-2) b^n \phi ) \right]
\nn\\ && \phantom{\frac{1}{2} \int_{M_d} |e|  +\ f}
\ - \ \frac{1}{2}(d-2) \left[\partial^n \phi
- \frac{1}{2}(d-2)b^n \phi \right]
\ - \ \frac{1}{2}(d-2) \o_n{}^{n \oplus} \phi
\Bigg) \ \ \Bigg\} \ .
\eea
The fields $\phi^n$ are auxiliary and are expressed in terms of
derivatives of $\phi$ by means of their equations of motion
\be
\phi_{n} = - \frac{2}{d-2}\ \partial_{n} \phi \ + \ b_n \phi \, ,
\ee
which leads to the equivalent action
\bea
\label{Aa2}
S & = & \left( \frac{a}{d-2} - \frac{g}{2} \right) \int_{M_d} |e|
\left[\ \partial_{n} \phi \ \partial^{n} \phi
\ - \ \frac{(d-2)}{4(d-1)}\, R(e)\, \phi^2
\ \right] \ .
\eea
In order to reach~(\ref{Aa1}) and~(\ref{Aa2}) we have performed
d-dimensional partial integrations. The action (\ref{Aa2}) differs
from (\ref{a2}) by an overall factor only.

\end{document}